\pdfoutput=1
\documentclass[11pt,letterpaper]{article}
\usepackage{jcappub}
\usepackage{graphicx,subfigure}
\usepackage{relsize}

\newcommand{\beq}{\begin{eqnarray}}
\newcommand{\eeq}{\end{eqnarray}}
\newcommand{\dP}{\delta\hspace{-.3mm}{\cal P}}
\newcommand{\teq}{t_{\rm eq}}

\newcommand{\Meq}{M_{\rm eq}}

\begin{document}

\title{Early structure formation from cosmic string loops}

\author[a]{Benjamin Shlaer}
\author[a]{Alexander Vilenkin}
\author[b]{Abraham Loeb}
\affiliation[a]{Institute of Cosmology, Department of Physics \&
Astronomy, Tufts University, \\ 212 College Avenue, Medford, MA 02155, USA}
\affiliation[b]{Institute for Theory \& Computation, Harvard-Smithsonian Center for
Astrophysics, \\60 Garden Street, Cambridge, MA, 02138, USA}
\emailAdd{shlaer@cosmos.phy.tufts.edu, vilenkin@cosmos.phy.tufts.edu, aloeb@cfa.harvard.edu}

\abstract{
We examine the effects of cosmic strings on structure formation and on the ionization history of the universe.  While Gaussian perturbations from inflation are known to provide the dominant contribution to the large scale structure of the universe, density perturbations due to strings are highly non-Gaussian and can produce nonlinear structures at very early times.  This could lead to early star formation and reionization of the universe.  We improve on earlier studies of these effects by accounting for high loop velocities and for the filamentary shape of the resulting halos.  
We find that for string energy scales $G\mu \gtrsim 10^{-7}$, the effect of strings on the CMB temperature and polarization power spectra can be significant and is likely to be detectable by the Planck satellite.
We mention shortcomings of the standard cosmological model of galaxy
formation which may be remedied with the addition of cosmic strings, and comment on other possible observational implications of early structure formation by strings.
}

\maketitle
\flushbottom

\section{Introduction}

Cosmic strings are linear topological defects that could be formed at a phase transition in the early universe \cite{Kibble}.  They are predicted in a wide class of particle physics models.  Some superstring-inspired models suggest that fundamental strings may also have astronomical dimensions and play the role of cosmic strings \cite{Sarangi:2002yt,Dvali,Copeland}.  

Strings can be detected through a variety of observational effects.  Oscillating loops of string emit gravitational waves -- both bursts and a stochastic background.  They can also be sources of ultrahigh-energy cosmic rays.  Long strings can act as gravitational lenses \cite{Vlensing,STlensing,SWlensing} and can produce characteristic signatures in the CMB: discontinuous temperature jumps \cite{Gott,Kaiser} along the strings (Gott-Kaiser-Stebbins effect) and a $B$-mode polarization pattern due to rotational perturbations induced by strings \cite{Seljak,PogosianWyman,Bevis:2007qz}.   Formation and evolution of cosmic strings and their possible observational effects have been extensively discussed in the literature; for a review and further references see, e.g., \cite{Vilenkin-Shellard,CopelandKibble,CPV}.

The strength of gravitational interactions of strings is characterized by the dimensionless number $G\mu$, where $\mu$ is the mass per unit length of string and $G$ is Newton's constant.  
Early work on cosmic strings was largely motivated by the idea that oscillating string loops and wakes formed by rapidly moving long strings could serve as seeds for structure formation \cite{Zeldovich,AV81,SilkV}.  
This scenario, which required $G\mu\sim 10^{-6}$, has been conclusively ruled out by CMB observations.  Present CMB bounds constrain $G\mu$ to be less than about $2\times 10^{-7}$ \cite{Pogosian,Fraisse,Pogosian:2008am,Battye,Landriau,Dvorkin}.  A much stronger bound, $G\mu < 4\times 10^{-9}$, has recently been claimed in \cite{vanHaasteren:2011ni}  (but see also \cite{Sanidas:2012ee}) and \cite{Demorest:2012bv} based on the millisecond pulsar observations of the gravitational wave background.  (We shall argue in section \ref{sec:discussion} that this bound could be significantly relaxed, in the light of the latest string simulations.)  

The CMB and other data are consistent with a Gaussian spectrum of density perturbations predicted by the theory of inflation; the contribution of strings, if any, can account for no more than $2\%$ of the spectral power.  This does not mean, however, that strings always played a
subdominant role in structure formation.  Density perturbations due to strings are highly non-Gaussian and can produce nonlinear structures at very early times.  This could result in early star formation and reionization of the universe \cite{Rees,Hara,Avelino,PV,OV,Hernandez:2011ym}.  Such early structure formation could manifest itself in CMB temperature and polarization \cite{Brandenberger3} spectra and could also produce a detectable 21-cm signal \cite{Khatri,Brandenberger,Brandenberger2,Pogosian3}.

These and other effects of strings depend on the details of string evolution, which until recently remained rather uncertain.  Early string simulations suggested that loops produced by the string network are short-lived and/or very small\footnote
{The simulation reported in \cite{Ringeval1} found a scaling loop population including large but fragmenting loops. Subsequent analytic work \cite{Ringeval2} assumed small loop production to be more significant for the loop energy density.
}
\cite{Ringeval1,Ringeval2,Bennett,Allen,Martins}.  Hence, it was assumed in Refs.~\cite{Rees,Hara,Avelino,PV,Brandenberger3,Khatri,Brandenberger,Brandenberger2} that the main effect on structure formation was due to long string wakes.  However more recent high-resolution simulations have demonstrated that a substantial\footnote{The discrepancy between this result and the results of \cite{Ringeval2} will be addressed in a future publication.} fraction of the network energy goes into large non-self-intersecting loops, having length of about $5\%$ of the horizon \cite{OVV,OVan,BOS}.   This pattern is established only after a long transient period dominated by very small loops; apparently it is this period that was observed in previous simulations.  
Ref.~\cite{OV} studied structure formation and reionization in this new string evolution scenario, 
but some important aspects of that scenario had not yet been recognized at the time.

In the present paper we shall reexamine early structure formation by cosmic strings and its effect on the ionization history of the universe.  An important fact not taken into account in Ref.~\cite{OV} is that string loops are typically produced with high velocities: $v\sim 0.3$ for the largest loops and even higher for smaller ones \cite{BOS}.  Accretion on such rapidly moving loops is rather different from spherical accretion on a stationary point mass which was assumed in \cite{OV}.  We also make use of the latest string simulations which yielded more reliable results for the size and velocity distributions of cosmic string loops.

The paper is organized as follows.  In the next section we review the evolution of cosmic strings and introduce the relevant loop distributions.  In section \ref{sec:accretion} we study accretion of matter onto a moving loop, first treating the loop as a point mass and then accounting for its finite size (which turns out to be significant).  We find that loop seeded halos have the form of highly elongated filaments which then fragment into smaller ``beads".  The resulting halo spectrum is calculated in section \ref{sec:spectrum}.  The baryon collapse fraction in string-seeded halos is calculated in section \ref{sec:collapse} and is then used in section \ref{sec:ionization} to study the reionization history of the universe and the possible effect of strings on the CMB temperature and polarization spectra.
Finally, in section \ref{sec:discussion} we comment on possible observational implications of the early structure formation by cosmic strings.  In appendix \ref{sec:rocket} we discuss the ``rocket effect" -- the self-acceleration of string loops due to asymmetric emission of gravitational waves -- and show that it is not significant for structure formation.

\section{Cosmic string evolution}\label{sec:evolution}

\subsection{Long strings}

An evolving network of cosmic strings consists of two components: long strings and sub-horizon closed loops.  The long string component evolves in a scaling regime in which the typical distance between the strings and all other linear measures of the network remain at a constant fraction of the horizon size $d_{\rm h}(t)$.  This dynamic is maintained by the production of loops via reconnection at string intersections, and subsequent evaporation of loops by gravitational radiation.  On super-horizon length scales, long strings have the form of random walks, while on smaller scales they exhibit a wide spectrum of wiggles and kinks, which are remnants of 
earlier reconnections down to the scale of initial formation.
These small-scale features are gradually smoothed out by expansion, small loop production, and by gravitational radiation back-reaction.

The total energy density in the long (``infinite") strings can be expressed as 
\beq
\rho_\infty = \mu/(\gamma d_{\rm h})^2,
\label{rhoinf}
\eeq
where $\mu$ is the mass per unit length of string (equal to the string tension), the radiation era horizon distance is $d_{\rm h} = 2t$, $\gamma$ is a constant coefficient, and the quantity $\gamma d_{\rm h}$ defines the average inter-string distance.   The idea that the network properties are determined by the horizon distance alone is known as the 
scaling hypothesis \cite{Kibble,Vilenkin:1981iu}; for the long string component it has been confirmed through numerical simulations \cite{Albrecht:1984xv,Bennett,Allen}.

The coefficient $\gamma$ in (\ref{rhoinf}) depends on the reconnection probability $p_{\rm rec}$ of intersecting strings.  For "ordinary" particle physics strings, $p_{\rm rec}=1$, and simulations give $\gamma \approx 0.15$ in the radiation era.  For cosmic superstrings, the reconnection probability is expected to be much smaller, $0.1\gtrsim p_{\rm rec}\gtrsim 10^{-3}$ \cite{JJP}, resulting in a smaller value of $\gamma$ and a denser string network.  Simple arguments suggest that $\gamma\propto p_{\rm rec}^{1/2}$ \cite{Damour,Sakellariadou}, but simulations indicate a weaker dependence \cite{Shellard}.  Here we shall focus on ordinary strings\footnote{We assume that strings are accurately described by the Nambu-Goto equations of motion.  Abelian field theory simulations (see \cite{Hindmarsh:2008dw} and references therein) yield different network properties, suggesting that the main energy loss mechanism of the network is direct production of massive particles from long strings.  However, analysis in Refs.~\cite{ShellardMoore,OlumJose} indicates that massive particle production is a transient phenomenon, caused by short-wavelength excitations in the initial state.}
with $p_{\rm rec}=1$.  

\subsection{Loop distribution: a simple model}
\label{subsec:simple-n}

The length distribution of loops produced by the network (the so-called loop production function) obtained in recent simulations has a double-peak structure, indicating two different populations of loops.  First, there is a scaling loop distribution, with a typical loop length of about $5\%$ of the horizon and a wide tail extending to smaller scales.  Then there is a non-scaling distribution of very small loops with sizes comparable to the initial scale of the network at formation.  The hight of the non-scaling peak is observed to decrease somewhat in the course of the simulation, with the extra power contributing to the short-length tail of the scaling peak.  One might expect that the non-scaling peak will eventually disappear \cite{OVV,OVan,BOS}, but 
Ref.~\cite{Polchinski} argued that it could also survive at late times.  In the latter case, the asymptotic double-peak loop distribution will still scale, but the typical loop size in the short-length peak will be set by the gravitational radiation damping.  

At present one can only guess which of the two options will be supported by future simulations, but fortunately this is not important for the purposes of the present paper.  The non-scaling loops are highly relativistic, with the dominant part of their energy being kinetic energy.  This energy redshifts with the expansion and has very little effect on the gravitational clustering.  The same applies to loops in the tail of the scaling distribution.  Only loops near the peak of the scaling distribution, which are formed with mildly relativistic velocities, are relevant for structure formation.
Moreover, halos formed by small loops at the tail of the distribution have small virial velocities.  This inhibits cooling and star formation in such halos.

The string loops of interest to us here will be those which formed during the radiation era but have not yet decayed at $t_{\rm eq}$.  The energy density of loops that were chopped off the network in one Hubble time is comparable to the energy density of long strings.  However, the loop energy redshifts like matter, while the long string energy redshifts like radiation.  So, if loops are long and live much longer than a Hubble time, they dominate the energy of the network and play dominant role in structure formation.

Using energy conservation, the power flowing into loops per unit physical volume
during the radiation era must obey
\beq\label{energy-balance}
{\dot{\rho}}_{\to {\rm loops}} = \frac{1 - \langle v_\infty^2\rangle}{t}\rho_\infty = \frac{\cal P \mu}{d_h^3},
\eeq
where $\langle v^2_\infty\rangle$ is the mean square velocity of the long strings.
Simulations give $\langle v_\infty^2\rangle \approx 0.4$ and ${\cal P} := \mu^{-1}d_h^3{\dot \rho}_{\to {\rm loops}} \approx 50$ in this epoch.
Neglecting  (for the time being) the center-of-mass loop velocity, the comoving number density distribution of loops $n(t,m)$ should satisfy
\beq\label{eq:loop-power-continuity}
\frac{1}{a^3(t)}\frac{d}{dt}\left( \int_0^\infty n(t,m)m dm\right) =  \dot{\rho}_{\to loops}.
\eeq
Here $a(t)$ is the scale factor, and the comoving loop distribution is related to the physical distribution by $n(t,m) = a^3(t) n_{\rm phys}(t,m)$.

Since we are only interested in loops near the peak of scaling distribution, it is a reasonable approximation to assume that all relevant loops are formed having length equal to a fixed fraction of the horizon size.  Hence, we set the loop mass at formation to be
\beq
m = \alpha\mu d_{\rm h},
\eeq
with $\alpha \approx 0.05$.  The distribution of such loops obeys
\beq
\frac{1}{a^3(t)}\frac{d}{dt}\left[n(t,m)m\right] = \delta(m -2\mu \alpha t)\frac{\dP\mu}{8t^3},
\eeq
and so
\beq
n(t,m)= \frac{a^3(\frac{m}{2\mu\alpha})}{m}\frac{\dP\mu}{8\left(\frac{m}{2\mu\alpha}\right)^3 2\mu\alpha}=
\frac{\sqrt{\alpha}\mu^{3/2}\dP}
        {4\sqrt{2}m^{5/2} t_{eq}^{3/2}}\qquad {\mbox {for }} m \leq 2\mu\alpha t, \label{rad-loop-density}
\eeq
and $n(t,m)=0$ otherwise.
Here, $\dP\approx 7$ reflects the fact that the
effective power flowing into these large loops is substantially less than the total power ${\cal P}\approx 51$ suggested by simulations \cite{BOS}.  We have also used $a(t)=(t/t_{eq})^{1/2}$, normalizing to $a(t_{eq})=1$.

The apparent divergence
of loop mass density $m\,n(t,m)dm$ at small $m$ is absent if we include the decay of loops due to gravitational radiation, $\dot{m} \approx -\Gamma G\mu^2$, with $\Gamma \approx 50$.
This correction is only significant for very small loops.

\subsection{Incorporating loop speed}\label{sec:speeds}

We now perform a more precise calculation which includes finite velocity effects, which are significant. 
We characterize the number density of loops by their mass $m$ and speed $v$.  All quantities will be expressed
per comoving volume from here onward.
The comoving number density of cosmic string loops with mass between $m$ and $m + dm$ 
and with speed between $v$ and $v + dv$ is given by $n(t,m,v)dv\,dm$.  The loop production function is
 $g(t,m,v)dt\,dv\,dm$, the number of such loops per comoving volume produced by the cosmic string network between time $t$ and $t + dt$.
The redshifting of loop speeds is given by $\dot{v} \approx - H v$ where $H = \dot{a}/a = 1/(2t)$ is the Hubble rate, and we are assuming non-relativistic $v$, 
which is justified for the subset of loops which contribute to star formation.

We can integrate the production rate to find the number density per comoving volume
\beq
n(t,m,v) &=& \int_0^t dt' g(t',m',v') \frac{\partial m'}{\partial m}\frac{\partial v'}{\partial v}.
\eeq
This states that the number of loops of mass $m$ and speed $v$ at time $t$ is the integral of the production rate of loops over all prior times $t' \leq t$, 
where the relevant production is of loops which will eventually have mass $m$ and speed $v$.  Hence the first step is to write down the solution to the flow, namely
\beq
m'(t';m,t) &\approx& m + \Gamma G\mu^2(t - t'),\label{eq:m-flow}\\
v'(t';v,t) &\approx& v\frac{a(t)}{a(t')}\label{eq:v-flow}.
\eeq
These tell us e.g., what speed $v'$ a loop must have at production time $t'$ in order for it to have a speed $v$ at time $t$.
The Jacobian factor $\frac{\partial v'}{\partial v}$ captures the changing size of the volume element $dv$.  The same applies to $m$ as well.  We are neglecting the effects on $v$ of
anisotropy in the loop gravitational radiation, also known as the ``rocket effect", which causes the loops to accelerate as they evaporate.  This will be justified in the Appendix.

The energy of a loop is given by $m/\sqrt{1 - v^2}$, and so using (\ref{energy-balance}),
\beq
\frac{{\cal P}\mu}{d_{\rm h}^3(t)} = \int_0^\infty\int_0^1 \frac{m}{\sqrt{1-v^2}}\frac{g(t,m,v)}{a^3(t)}dv\,dm.
\eeq
Since all quantities of interest are linear in $g$, we can maintain generality while treating $g$ as a delta-function source, i.e.,
\beq\label{eq:delta-form}
g(t,m,v) = \frac{a^3(t)\dP\mu\sqrt{1-v^2}}{d_{\rm h}^3(t)m} \delta\left(m- \alpha\mu d_{\rm h}(t) \right) \delta\left(v - \alpha_v \right),
\eeq
where in this form we can imagine taking the values suggested by simulations, namely 
\beq\label{eq:delta-approx}
\alpha = 0.05,\quad \alpha_v = 0.3, \quad \dP = 7.  
\eeq
The results using the actual loop production function, including small relativistic loops, can always be obtained by integrating any expression involving $\dP$ with the replacement $\dP \to $(numerical loop-production distribution) using simulation data from e.g., \cite{BOS}.  Because of this substitution possibility, we will refer to (\ref{eq:delta-form}) as the delta-function form of $g$, and (\ref{eq:delta-approx}) as the delta-function approximation for $g$.  

Combining the equations of this section, the comoving loop number density is
\beq
n(t,m,v) 
&=& \int_0^t dt' g(t',m',v') \frac{\partial m'}{\partial m}\frac{\partial v'}{\partial v},\\ \nonumber
&=& \int_0^t dt' \frac{a^3(t')\dP\mu}{d_{\rm h}^3(t')m'} \delta\left(m'- \alpha\mu d_{\rm h}(t') \right) \delta\left(v' - \alpha_v \right)\frac{a(t)}{a(t')},\\ \nonumber
&\approx& \frac{\sqrt{\alpha}\dP\mu^{3/2} \delta\left(v - \sqrt{\tfrac{m}{2\alpha\mu t}}\alpha_v\right)}
{4\sqrt{2}\left(m+\Gamma G\mu^2 t\right)^{5/2}t_{\rm eq}^{3/2}}.
\eeq

\section{Accretion onto a cosmic string loop}\label{sec:accretion}

As we explained in Section \ref{subsec:simple-n}, the dominant loop energy density will always be from loops produced in the radiation era.  This is because
the integrated power flowing into loops scales like $1/t^2$, whereas the subsequent dilution only scales like $1/a^3$.  Thus in the radiation-era (when $a\sim t^{1/2}$), loops pile up 
from early times.  This growth is cut off by loop evaporation to gravitational radiation, which is a slow process for low tension strings.  So structure formation
is mainly sensitive to radiation-era loops which have survived to the time of matter domination.  

Of particular interest are halos which become large enough for stars to form \cite{Loeb-book}.  This requires the baryons to collapse to sufficient density, which can only happen
due to dissipation.  When the neutral hydrogen atoms have large enough virial velocity, collisions will have sufficient center-of-mass energy to excite their electrons and
radiate.  The energy escaping with the radiated photons ensures that the hydrogen loses gravitational potential energy.  This cooling moves the hydrogen toward the center of the halo.  The critical virial temperature for efficient cooling $T_\ast$ can be as low as 200K in the case of molecular hydrogen, but we will neglect this mechanism  since the molecule is too fragile to survive UV light from the very first stars.  Instead, we will consider two possible values, 10$^4$K and  $10^3{\rm K}$, below which star formation does not occur.\footnote{Because of the relative motion between the baryon and dark matter fluid, smaller halos will not lead to significant star formation \cite{Stacy:2010gg,Tseliakhovich:2010yw}.}  Because the virial temperature is proportional to a positive power of the halo mass, which is proportional to the seed (loop) mass, there is a minimum loop size which can lead to star formation.  We
can conclude that early stars can only come from loops produced late enough to have this size.  We will denote the minimum scale factor after which star-forming loops can be produced by $a^\ast_i$, which we will now estimate.  Throughout, we define the scale factor at matter-radiation equality to be unity, $a(t_{\rm eq}) = 1$.

\subsection{Spherical accretion}

For a spherical halo of mass $M$ formed at redshift $z$, the
virial temperature is given by \cite{Loeb-book}
\beq\label{eq:Tvir}
T_{\rm vir} = 10^4 \left(\frac{M}{10^8 M_\odot}\right)^{2/3}\left(\frac{1+z }{10}\right){\rm K},
\eeq
and so the minimal mass of a halo that has $T_{\rm vir} \geq 10^4~{\rm K}$ at redshift $z$ is
\beq
M_\ast (z)=  3\times 10^9 M_\odot(1+z)^{-3/2}\left(\frac{T_\ast}{10^4{\rm K}}\right)^{3/2}.
\label{M*}
\eeq

 Because the vast majority of loops are moving rather quickly, the more appropriate bound on production time considers elongated filaments, rather than spherical halos.  Elongated halo filaments have a virial temperature \cite{Eisenstein:1996pr} dependent entirely on their linear mass density $\mu_{\rm fil}$, which we will calculate below.  These filaments
 will later collapse into beads, which subsequently merge into larger beads, but this process does not significantly affect the virial temperature. 

\subsection{Accretion onto a moving point mass}

If we assume the loop is non-relativistic and has velocity $v_{\rm eq}>0$ in the $+y$-direction at $t_{\rm eq}$, its subsequent velocity is given by
\beq\label{eq:vloop}
v(a) = \frac{v_{\rm eq}}{a},
\label{v}
\eeq
where $a = (t/t_{\rm eq})^{2/3}$.
The trajectory in comoving coordinates is then 
\beq\label{eq:trajectory}
y(a) =  3 v_{\rm eq} t_{\rm eq}\left(1-\frac{1}{\sqrt{a}}\right).
\eeq
Using our delta-function loop production function, the loop velocity at equality is
\beq
v_{\rm eq} \approx \alpha_v a_i , 
\eeq
where $a_i<1$ is the scale factor at loop production, relative to $a_{\rm eq} = 1$. 

Given a point mass in the above trajectory, we can find the cylindrically symmetric turn-around surface by considering a particle at comoving initial location $(x_0,y_0,0)$.  (Here we closely follow the analysis in \cite{Bertschinger}.)
Let us label the moment of closest approach of the mass by the scale factor $a_0$, i.e., the loop trajectory $y(a)$ obeys $y(a_0) = y_0$.  Using the impulse approximation, the velocity kick on the particle due to the passing point mass is
\beq
v_x \sim \frac{Gm}{(a_0x_0)^2}\Delta t_0 \sim \frac{Gm}{a_0x_0v(a_0)},
\eeq
where
\beq
\Delta t_0 \sim \frac{a_0x_0}{v(a_0)}.
\eeq
This particle will then be displaced after one subsequent Hubble time $H_0^{-1} \sim t_{\rm eq} a_0^{3/2}$ by an amount
\beq
a_0\Delta x_0 \sim v_x t_{\rm eq} a_0^{3/2},
\eeq
and so the corresponding density perturbation is
\beq
\delta_0 \sim \frac{\Delta x_0}{x_0} \sim \frac{Gmt_{\rm eq}}{x_0^2 v(a_0)a_0^{1/2}},
\eeq
which grows to be
\beq
\delta(a) \sim \delta_0\frac{a}{a_0} \sim  \frac{Gmt_{\rm eq}a}{x_0^2 v(a_0)a_0^{3/2}}.
\eeq
The turnaround surface is given by $\delta(a) \sim 1$, so the profile of the collapsed region from a passing point mass with arbitrary velocity $v(a_0)$ is
\beq\label{eq:turnaround}
x_{\rm ta}^2(a_0,a) \sim  \frac{Gmt_{\rm eq}a}{ v(a_0)a_0^{3/2}},
\eeq
where the time dependence is in $a = (t/t_{\rm eq})^{2/3}$, and the $y$-dependence is in $a_0(y_0)$ via (\ref{eq:trajectory}) with $y(a_0) = y_0$.  Notice $a \geq a_0$, since the turnaround surface extends only behind the loop's $y$-position.
The halo mass is then given by the total mass inside the turnaround surface,
\beq
M(t) &=& \rho a^3(t)\int_0^{y(a(t))}\pi x_{\rm ta}^2(a_0(y_0),a(t))dy_0 \,=\, \rho a^3(t)\int_{t_{\rm eq}}^t\pi x^2_{\rm ta}\frac{d y(a(t_0))}{d t_0}dt_0\\ \nonumber
&=&\rho a^3(t)\int_{t_{\rm eq}}^t\frac{\pi x^2_{\rm ta}v(a_0)}{a_0} dt_0\,=\, \frac{1}{6\pi G t_{\rm eq}^2}\int_{t_{\rm eq}}^t\frac{\pi Gmt_{\rm eq}a(t)}{a_0^{5/2}}dt_0
\,=\, \frac{m a(t)}{6}\int_{t_{\rm eq}}^{t}\frac{t_{\rm eq}^{2/3}}{t^{5/3}_0}dt_0\\ \nonumber
&=& \frac{m}{4}\left[a(t)-1\right]. \label{eq:point-mass-halo}
\eeq
Interestingly, the loop trajectory does not affect the halo mass at this level of approximation.
A more accurate analysis by Bertschinger \cite{Bertschinger}, using the Zel`dovich approximation, gives $M(t) = \frac{3}{5} m a(t)$.  We will simply use 
\beq
M(t) \sim m a(t).
\eeq

We can solve for the shape and size of the turnaround surface by combining (\ref{eq:vloop}), (\ref{eq:trajectory}) \& (\ref{eq:turnaround}).
This gives
\beq\label{eq:profile}
x_{\rm ta}(y,a) = t_{\rm eq}\sqrt{\frac{2\alpha G\mu a_i a}{\alpha_v} \left(1-\frac{y}{3\alpha_va_it_{\rm eq}}\right)}\qquad {\rm for} \quad 0\leq y \leq 3\alpha_v a_i t_{\rm eq}\left(1-\frac{1}{\sqrt{a}}\right).
\eeq

Notice that these halos are very elongated:  
the eccentricity after a Hubble time is given by
\beq
\left.\frac{y}{x_{\rm ta}}\right|_{a \sim 2} \approx \frac{\alpha_v^{3/2}a_i^{1/2}}{\sqrt{\alpha G\mu}} \sim 10^3 \mu_{-8}^{-1/2}.
\eeq
Here we have assumed $a_i = \sqrt{t_i/t_{\rm eq}}\sim 0.1$, and used the delta-function approximation $\alpha_v \sim 0.3$ and $\alpha \sim 0.05$.
We refer to these elongated structures as filaments.  Halos will form from linear instabilities of the filaments.

In this discussion we disregarded the so-called rocket effect -- the self-acceleration of the loop due to asymmetric emission of gravitational waves.  This has negligible effect on the loop's velocity, except toward the end of the loop's life, when the loop can be accelerated to a mildly relativistic speed, $v\lesssim 0.1$.  The loop trajectory with the rocket effect included is discussed in the Appendix, where it is shown that the rocket acceleration has little influence on halo formation.

\subsection{Accretion onto a finitely extended loop}

Since the loop has a finite radius $R = \beta m/\mu$, with $\beta \sim 0.1$, only the matter outside this distance will feel a momentum kick from the passing loop.\footnote{The rapidly oscillating string will leave wakes of overdensity behind the fast segments, i.e., even inside the loop radius $R$.  We neglect this effect, since only a small fraction $\sim 8\pi G\mu/v$ of the material is affected, and the wakes are probably too thin to allow star formation.}  We should then only consider the portion of the turnaround surface where
\beq
x_{\rm ta} > R/a_0,
\eeq
or using (\ref{v}) \& (\ref{eq:turnaround}),
\beq\label{turnaround}
\alpha_v\frac{a_i}{a_0} 
\, <\,    \frac{G\mu a a_0^{1/2}}{2\beta^2\alpha a_i^2},
\eeq
where again, $a_0$ should be thought of as a measure of the $y$-coordinate given by $y_0 = y(a_0)$ above.
This equation is a restriction on the validity of the turnaround surface (\ref{eq:profile});  in places where the turnaround surface is smaller than the comoving loop radius, it does not exist and so should be thought of as ending rather than closing.  

For the faster loops (whose turnaround surfaces have smaller physical radius c.f. (\ref{eq:turnaround})), the loop radius will entirely cloak the turnaround surface at the onset of matter domination.  Because the turnaround surface radius grows relative to the comoving loop radius, there will eventually be a time when the surface emerges.  The surface will emerge as a hoop surrounding the loop, which grows into a cylinder.  The front of the cylinder will move forward with the loop, and continue to grow in diameter with expansion.  The back of the cylinder will extend back to the location where the turnaround surface is smaller than the comoving loop radius.  
Eventually the back of the cylinder will reach $y=0$, i.e., the location of the loop at the onset of matter domination.

We can now talk of three distinct phases of loop-seeded filaments in the matter era.  The early type are those which are not accreting, because no part of the turnaround surface has emerged from behind the loop radius.  The late type are filaments whose entire turnaround surface is larger than the loop radius.  Because the point-mass approximation holds in this case, the total filament mass is given by (\ref{eq:point-mass-halo}).  We will call these ``normal growth" filaments.  The intermediate type
of filament are those whose turnaround surfaces are growing both in the positive $y$-direction with the loop motion, as well as in the negative $y$-direction as more and more of the turnaround surface emerges from beneath the loop radius.  We will call these ``accelerated growth" filaments, since they are catching up from having zero mass  to eventually have normal mass.

\begin{figure}[htp]
\centering
\includegraphics[width=3.5in]{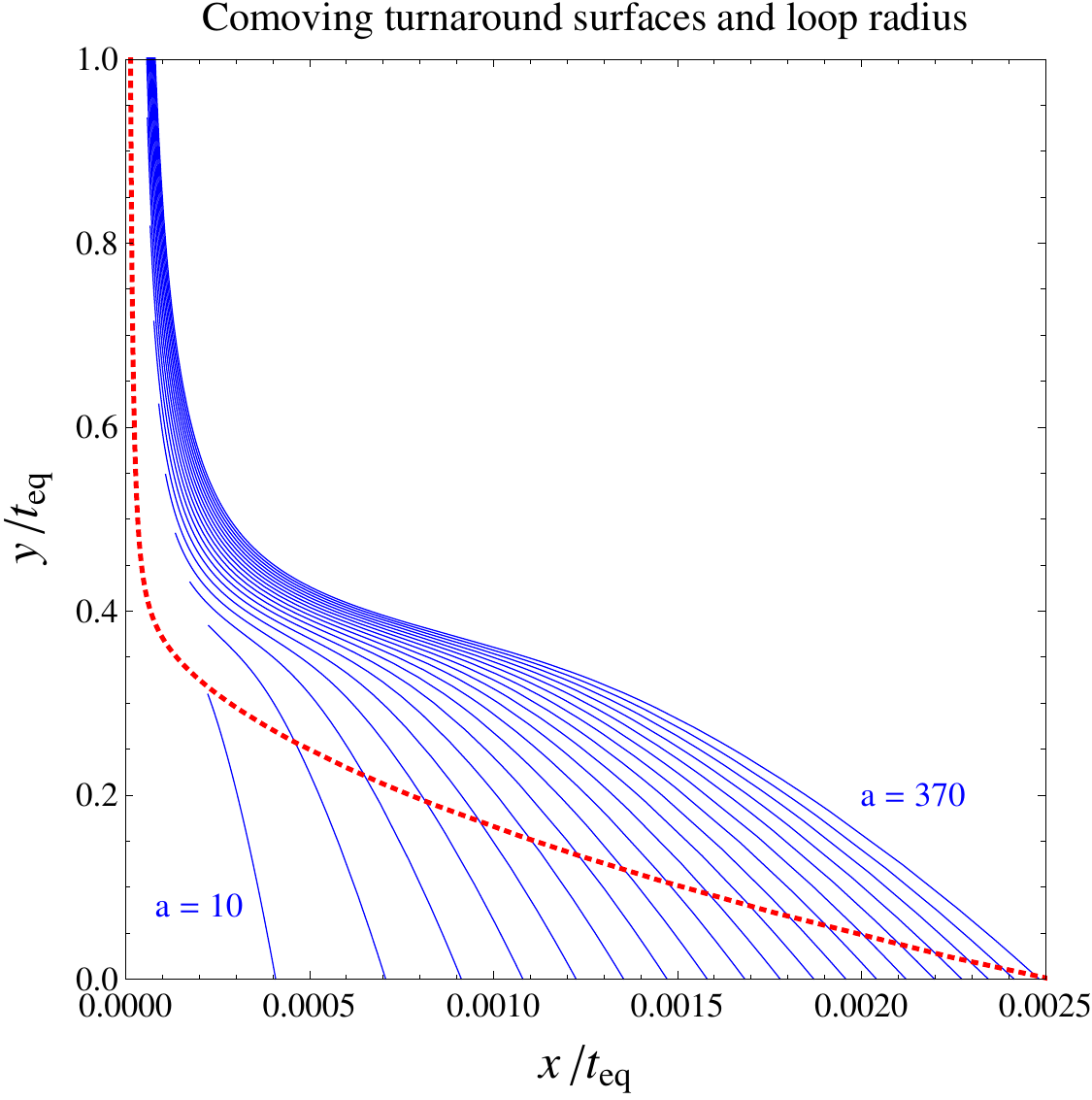}
\caption{Time-lapse ($\Delta a = 20$) illustration of comoving turnaround surfaces (thin blue lines) for a loop of tension $G\mu = 10^{-7}$.  The comoving loop extension is shown dotted in red.  Notice the aspect ratio is reduced by a factor of 400:1, and units are $t_{\rm eq}$.  One can read off $a_{\rm min} \approx 10,  a_{\rm max} \approx 370$ from the figure.  We use $a_i = 1/2$ and the delta-function approximation parameters of (\ref{eq:delta-approx}) to determine the mass and speed of the loop.  The rocket effect is included for completeness (but see Appendix).}\label{fig:turnaround}
\end{figure}

We can find when the turnaround surface first emerges from behind the loop radius $R$ by setting $a_0 = a$ in (\ref{turnaround}) 
to find
\beq\label{eq:amin}
a > a_{\rm min} = \left(\frac{2\beta^2\alpha\alpha_va_i^3}{G\mu}\right)^{2/5}.
\eeq
The turnaround surface stops its accelerated growth when (\ref{turnaround}) is satisfied all the way back to $y_0 = 0$, i.e. $a_0 = 1$, and
so the normal-growth regime for the filament
takes over after
\beq\label{eq:amax}
a > a_{\rm max} = \left(\frac{2\beta^2\alpha\alpha_va_i^3}{G\mu}\right) = a_{\rm min}^{5/2}.
\eeq
The turnaround surfaces for $a = 10,20...370$ are shown in figure \ref{fig:turnaround}.

Given these turnaround surfaces, the total filament mass can be shown to be
\beq\label{eq:filament-mass}
M_{\rm fil}(t) \quad = \quad \rho_{\rm eq}\int_{y_0^{\rm min}(t)}^{y(t)}x(t)^2_{\rm ta}dy_0 \quad \sim
\quad\left\{
\begin{array}{ll}
0& \quad a < a_{\rm min}\\
m(\frac{a^{5/3}}{a_{\rm min}^{5/3}} -1)&\quad a_{\rm min}<a < a_{\rm max}\\
m(a - 1)&\quad  a_{\rm max}<a.
\end{array}\right.
\eeq
Note that in the delta-function approximation for $g$ where $a_{\rm min}^{5/3}\propto a_i^2 \propto m$, the filament mass in the accelerated growth regime is the same for all loops, independently of their mass $m$.

With the loop trajectory from (\ref{eq:trajectory}), the physical length of the filament is given by
\beq
L_{\rm fil} \quad \sim \quad \left\{
\begin{array}{ll}
0& \quad a < a_{\rm min}\\
3 \alpha_va_it_{\rm eq}(\frac{a^{4/3}}{a_{\rm min}^{5/6}} - \sqrt{a})&\quad a_{\rm min}<a < a_{\rm max}\\
3 \alpha_va_it_{\rm eq}(a - \sqrt{a})&\quad  a_{\rm max}<a,
\end{array}\right.
\eeq
where $a_{\rm max} = a_{\rm min}^{5/2}$.  This results in a linear mass density
\beq
\mu_{\rm fil} \sim  \frac{M_{\rm fil}}{L_{\rm fil}}  \sim  \frac{2\alpha\mu a_i}{3\alpha_v}
\eeq
for normal growth filaments.\footnote{Accelerated growth filaments will have a slightly smaller linear mass density.  The difference will not be important except at very high redshifts when there are no star forming halos in the normal growth regime.}
The corresponding virial temperature is found in  \cite{Eisenstein:1996pr} to be $T_{\rm vir} \approx \tfrac{1}{2} m_p G\mu_{\rm fil}$, where the proton mass $m_p = 1.1\times 10^{13}$ K. Hence
\beq\label{eq:fil-T-vir}
T_{\rm vir} &\sim& \frac{G m_p\sqrt{m\alpha\mu}}{3\sqrt{2} \alpha_v \teq^{1/2}} \\
&\sim&10^4 a_i\mu_{-8}\left(\frac{\alpha}{0.05}\right)\left(\frac{0.3}{\alpha_v}\right){\rm K}.
\eeq
Thus star formation occurs only for loops formed after
\beq
a_i^\ast \sim \frac{1}{\mu_{-8}}\left(\frac{0.05}{\alpha}\right)\left(\frac{\alpha_v}{0.3}\right)\left(\frac{T_\ast}{10^4{\rm K}}\right).
\eeq
Note that if $T_\ast = 10^4$K and the string tension obeys $G\mu \leq 10^{-8}$, star formation can only be attributed to loops in the low-velocity tail of the loop production function, which is highly
suppressed.  

If we restrict our attention to the ionization history of the universe, which is sensitive only to the total fraction of baryons in stars, we can neglect the subsequent
dynamics of the filaments, which will fragment into beads which subsequently merge.
This process will not significantly increase the virial temperature, since the filament is already in (2D) virial equilibrium, and the collapse into beads preserves
the total energy.

\subsection{Longitudinal filament collapse (or lack thereof)}

Although we will find a linear instability of filaments toward collapse into bead-like halos, we can rule out
the merging of the entire filament into a single large halo for all but the slowest loops.  Here we consider the longitudinal collapse mode of the entire filament.

To show such a filament will not collapse, we can disregard the finite size of the loop, since accounting for
the finite size can only further delay collapse.
Then within a Hubble time of $t_{\rm eq}$ the filament mass is roughly the mass of the loop,
\beq
M_{\rm fil} \sim \alpha \mu t_i .
\eeq
The length of the filament at this time is $L_{\rm fil} \sim v_{\rm eq} t_{\rm eq}$, and it grows with the scale factor as
\beq
L_{\rm fil} \sim v_{\rm eq} t_{\rm eq} a,
\eeq
where the loop velocity at $t_{\rm eq}$ is $v_{\rm eq} \sim \alpha_v a_i$.
The average mass per unit length is
\beq
\mu_{\rm fil}  \sim  \frac{M_{\rm fil}}{L_{\rm fil}}  \sim  \frac{\alpha\mu a_i}{\alpha_v}.
\eeq

Let us first disregard the collapse along the filament axis.  Then,
at $t > t_{\rm eq}$, both the mass of the filament and its length grow like
$(t/t_{\rm eq})^{2/3}$, so $\mu_{\rm fil}$ remains constant.

Now let us estimate the time scale of the longitudinal collapse.  The
characteristic longitudinal velocity that parts of the filament develop in a
Hubble time $t$ due to the filament's self-gravity is
\beq
v \sim  (G\mu_{\rm fil} / L_{\rm fil}) t.
\eeq
The collapse sets in when this becomes comparable to the Hubble velocity,
$v_H \sim L/t$.  This gives the following estimate for the redshift of the
collapse:
\beq
(1+z) \sim  \alpha G\mu \alpha_v^{-3} (t_{\rm eq}/t_i)^{1/2}(1+ z_{\rm eq}).
\eeq
With $\alpha \sim 0.05$, $\alpha_v \sim 0.3$, $G\mu\lesssim 10^{-7}$ and  $t_i/t_{\rm eq} \gtrsim 10^{-4}$, we get $1+ z < 1$, i.e., no collapse.
This means that the filaments from cosmic string loops remain elongated by a large factor.

\subsection{Fragmentation of filaments into beads}\label{sec:filament-collapse}

As the filaments grow in mass and thickness from radial infall of dark matter (and baryons after recombination), the longitudinal expansion
maintains their linear mass density at a constant (in time) value $\mu_{\rm fil}(y) = \rho_{\rm eq} x^2_{\rm ta}(y,a)/a$.  This overdense cylinder will be unstable to collapse
into beads on scales longer than the Jeans length.  The fastest growing instability was found in \cite{Quillen:2010yc} to have
a comoving wavelength $\lambda_{\rm J} \sim 4\pi x_{\rm vir}$, where the virial radius of a cylindrically symmetric self gravitating gas
is $x_{\rm vir} = x_{\rm ta}/\sqrt{e}\approx x_{\rm ta}$, and so the comoving length of the typical bead is comparable to the transverse dimension of the cylinder,
\beq
\Delta y \sim 4 x_{\rm ta}.
\eeq
Longer wavelength instabilities represent merging of such beads, and occur on scales up to an order of magnitude longer.
Since we characterize the beads by the size of the longest unstable mode, the length of the bead after merging is
\beq
L_{\rm bead} \approx 10a\Delta y \approx 20\pi t_{\rm eq}\sqrt{\frac{a^3 a_i\alpha G\mu}{\alpha_v}} \approx 4\times 10^{-5}\sqrt{a^3 a_i\mu_{-8}}\mbox{ Mpc}.
\eeq
The number of beads is then
\beq
\nu_{\rm beads} = \frac{L_{\rm fil}}{L_{\rm bead}} .
\eeq

In accelerated growth filaments, the bead mass is
\beq\label{eq:acc-bead-mass}
M = \frac{M_{\rm fil}}{\nu_{\rm beads}} \approx 30 \alpha ^{7/6}\beta^{-2/3} \sqrt{a_i} \Meq \left(\frac{a G\mu }{\alpha_v}\right)^{11/6},
\eeq
where $\Meq=t_{\rm eq}/G = 3.6\times 10^{17} M_\odot$ is (roughly) the mass contained in a horizon-sized sphere at $t_{\rm eq}$.  

The mass of each halo from normal-growth filaments can likewise be estimated to be
\beq\label{eq:norm-bead-mass}
M &=& \frac{M_{\rm fil}}{\nu_{\rm beads}} \approx 36\Meq\left(\frac{G\mu\alpha a_i a}{\alpha_v}\right)^{3/2}.
\eeq

By requiring $T_{\rm vir} \geq T_\ast$, we can set the lower bound on the relevant bead mass at time $a(t)$ to be
\beq
M^{\rm acc}_\ast \approx 50 a^{11/6} \Meq \sqrt{\frac{T_\ast}{m_p}} \left(\frac{\alpha (G\mu) ^2}{\alpha_v^2 \beta}\right)^{2/3}
\eeq
for beads in accelerated growth filaments and
\beq
M^{\rm norm}_\ast \approx 200 \Meq a^{3/2}\left(\frac{T_\ast}{m_p}\right)^{3/2}
\eeq
for beads in normal growth filaments.

\section{Halo spectrum}\label{sec:spectrum}

We are interested in comparing models with cosmic strings of tension $G\mu = 10^{-8}\mu_{-8}$ with the standard structure formation scenario.  The halo spectrum of the standard scenario is
well approximated by the Sheth-Tormen mass function \cite{Sheth:1999mn}.  The $\Lambda_{\rm CDM}$ cosmology we use assumes the values of WMAP+BAO+$H_0$ \cite{Komatsu:2010fb}, which for our purposes is just $z_{\rm eq} = 3232$ and $\Meq = 3.62\times 10^{17} M_\odot$.  Because we are interested in early star formation, we will assume the growth function $D(z)$ scales simply as $D(z) \propto (1+z)^{-1}$.

We can calculate the spectrum of halos using the continuity equation.  As we only consider radiation-era loops, there is no source term, since every loop already exists by $a = a_{\rm eq} = 1$, and subsequently is associated with one filament, which in turn is associated with some number $\nu_{\rm beads}$ of halos.

The spectrum of halos $N(M,a)$ at any redshift $z = \frac{z_{\rm eq+1}}{a} - 1$ is determined entirely from the spectrum of loops $n(a,m,v)$ at $a = a_{\rm eq}$ via
\beq\label{eq:Ndv}
N(a,M) = \int \nu_{\rm beads}(a,m,v_{\rm eq}) n(a_{\rm eq},m,v_{\rm eq})\frac{\partial m}{\partial M}dv_{\rm eq},
\eeq
where
\beq
\nu_{\rm beads}(a,m,v_{\rm eq}) = \left\{
\begin{array}{ll}
0.08 v_{\rm eq}^{7/6} \left(\frac{\Meq}{m}\right)^{5/6} \left(\frac{G\mu}{\beta}\right)^{2/3} a^{-1/6} & \quad a_{\rm min} < a < a_{\rm max}  \vspace{2mm}\\
0.08 \left(\frac{\Meq v_{\rm eq}^3}{ma}\right)^{1/2} & \quad a_{\rm max} < a,
\end{array}
\right.
\eeq
\beq
n(a_{\rm eq},m,v_{\rm eq}) &=& \frac{\delta{\cal P}\mu^{3/2}\sqrt{\alpha}\,\delta\hspace{-1mm}\left(v_{\rm eq} - \alpha_v\sqrt{\tfrac{m}{2\alpha\mu t_{\rm eq}}}\right)}
{2(2 t_{\rm eq})^{3/2}\left(m+\Gamma G\mu^2 t_{\rm eq}\right)^{5/2}},
\label{namv}
\eeq
and $m(a;M)$ is given by combining (\ref{eq:acc-bead-mass}) or (\ref{eq:norm-bead-mass}) with the substitution 
\beq
a_i = \left(\frac{m}{2\alpha\mu t_{\rm eq}}\right)^{1/2}.
\label{ai}
\eeq
The delta-function enables us to integrate (\ref{eq:Ndv}), yielding $v_{\rm eq} = \alpha_v (m/2\alpha\mu t_{\rm eq})^{1/2}$.  We can neglect
the correction proportional to $\Gamma$, since the affected loops are too small to seed halos with $T_{\rm vir} \geq T_\ast$.

The Jacobian $({\partial m}/{\partial M})$ can be found from the loop mass
\beq
m(M) = \left\{
\begin{array}{ll}
3.0 \times 10^{-6} \frac{M^4}{\Meq^3} \frac{G\mu}{\beta} \left(\frac{\beta \alpha_v^{2}}{\alpha (G \mu)^{2} a^2}\right)^{11/3}&\quad a_{\rm min}<a < a_{\rm max}\\ 
\vspace{-4mm}\\
1.7\times 10^{-2} \frac{\alpha_v^2 M}{\alpha G\mu a^2}\left(\frac{M}{\Meq}\right)^{1/3} &\quad  a_{\rm max}<a.
\label{mM}
\end{array}\right.
\eeq
Note that although $m(M)$ is continuous at the transition between the accelerated and normal growth regimes at $a=a_{max}$, the derivative $({\partial m}/{\partial M})$ is not, resulting in a discontinuity in the halo spectrum (\ref{eq:Ndv}).

We can write $a_{\rm min}$ and $a_{\rm max}$ from (\ref{eq:amax}) without reference to $a_i$ by using (\ref{ai}),
\beq
a_{\rm max} &=& \frac{\beta^2 \alpha_v}{\sqrt{2\alpha}(G\mu)^{5/2}} \left(\frac{m}{\Meq}\right)^{3/2},\label{eq:amax-m}\\
a_{\rm min} &=& a_{\rm max}^{2/5}.
\eeq
We can also express $a_{\rm max}$ in terms of the halo mass $M$ by substituting the lower expression in (\ref{mM}) with $a=a_{\rm max}$ into (\ref{eq:amax-m}) and then solving for $a_{\rm max}$.  This gives
\beq
a_{\rm max} = \frac{0.2 \alpha_v}{G\mu} \left(\frac{\beta M}{\alpha \Meq} \right)^{1/2} = \frac{0.014}{\mu_{-8}}\left(\frac{M}{M_\odot}\right)^{1/2}.
\eeq

The halo spectrum is then given by
\beq
N(a,M) = 
\left\{
\begin{array}{ll}
N_{\rm acc}(a,M) &\quad a < a_{\rm max}\\ 
\vspace{-1mm}\\
 N_{\rm norm}(a,M)&\quad  a > a_{\rm max},
\end{array}\right.
\eeq
where
\beq
N_{\rm acc}(a,M)&=& 2.4\times 10^8  \frac{\alpha_v\beta\delta{\cal P}}{G^3M^4} \left(\frac{\Meq} {M}\right)^4
\left[\frac{\alpha}{\alpha_v^2 \beta}(G\mu a)^2\right]^{19/3},\\
   N_{\rm norm}(a,M) &=& \frac{2.5\alpha\delta{\cal P}}{\alpha_v G^3 M^4} \left(\frac{M}{\Meq}\right)^{4/3} (G\mu a)^2.
\eeq

\begin{figure}[htbp] 
   \centering
   \mbox{\hspace{-.2in}\subfigure{
   \includegraphics[width=3.1in]{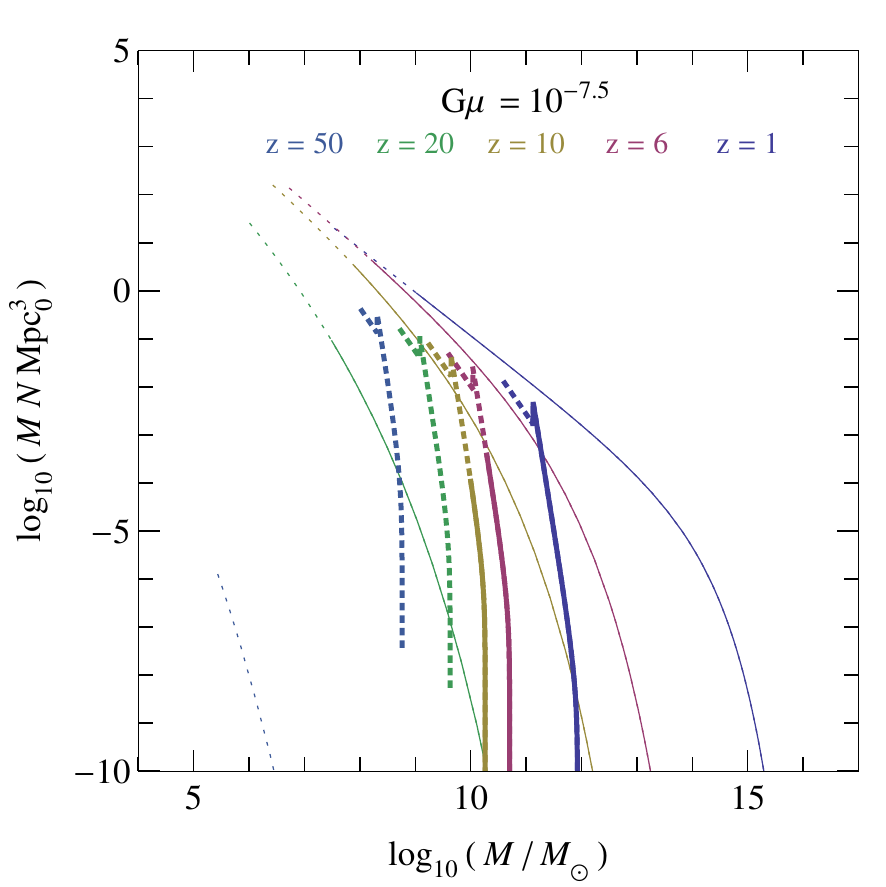} 
   }\subfigure{
     \includegraphics[width=3.1in]{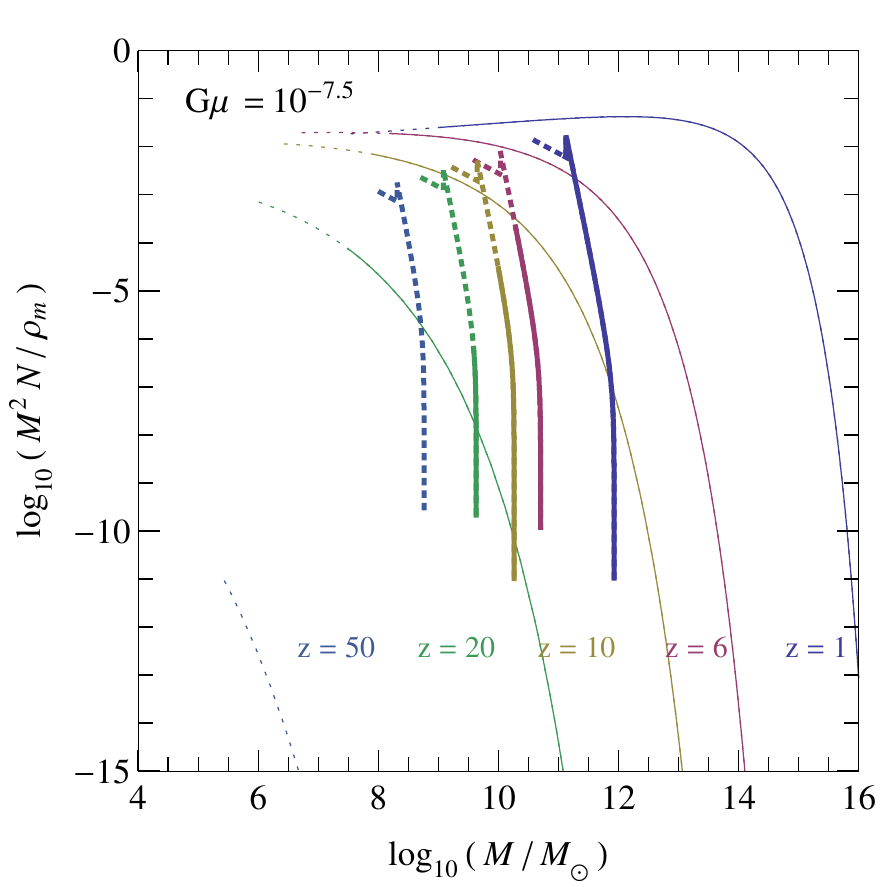}
    }}
   \caption{Number and mass densities of halos per logarithmic mass bin.  Solid lines indicate halos with $T_{\rm vir} > 10^4$K, and dotted lines extend this down to $10^3$K.    The
   Sheth-Tormen mass function is used for comparison (thin lines).   From right to left the redshifts are (1,6,10,20,50).  We use the delta-function approximation for $g$ with $\alpha = 0.05$, $\alpha_v = 0.3$, and $\delta{\cal P} = 7$.} 
   \label{fig:st75}
\end{figure}

\begin{figure}[htbp] 
 \centering
   \mbox{\hspace{-.2in}
      \subfigure{\includegraphics[width=3.1in]{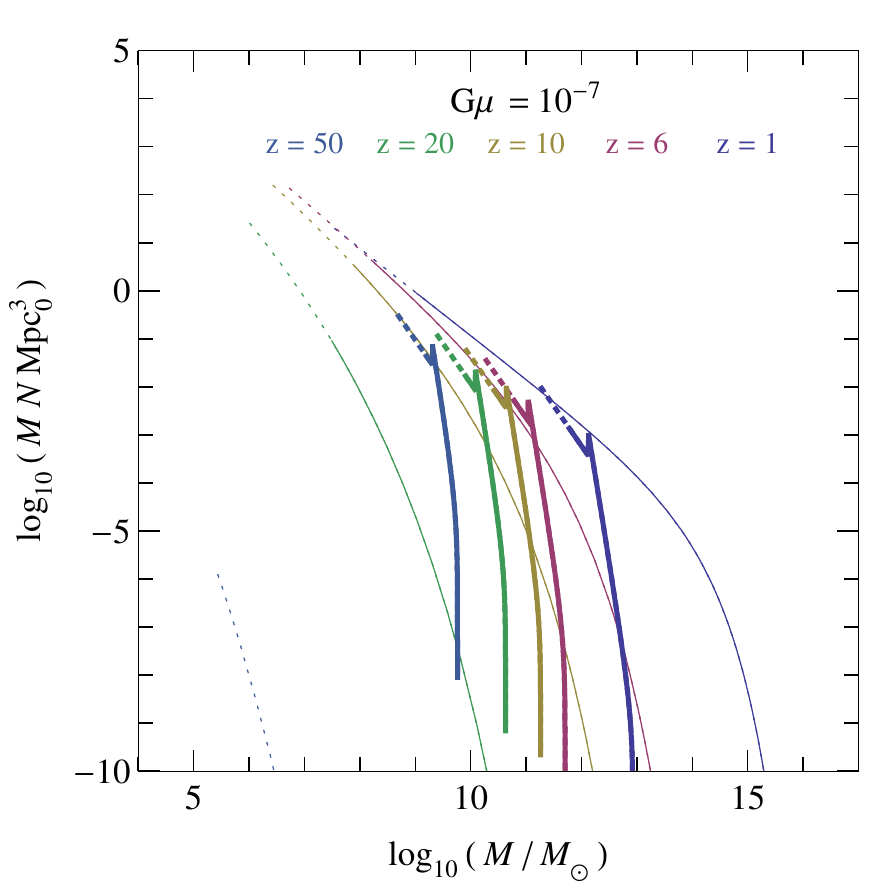}}
      \subfigure{\includegraphics[width=3.1in]{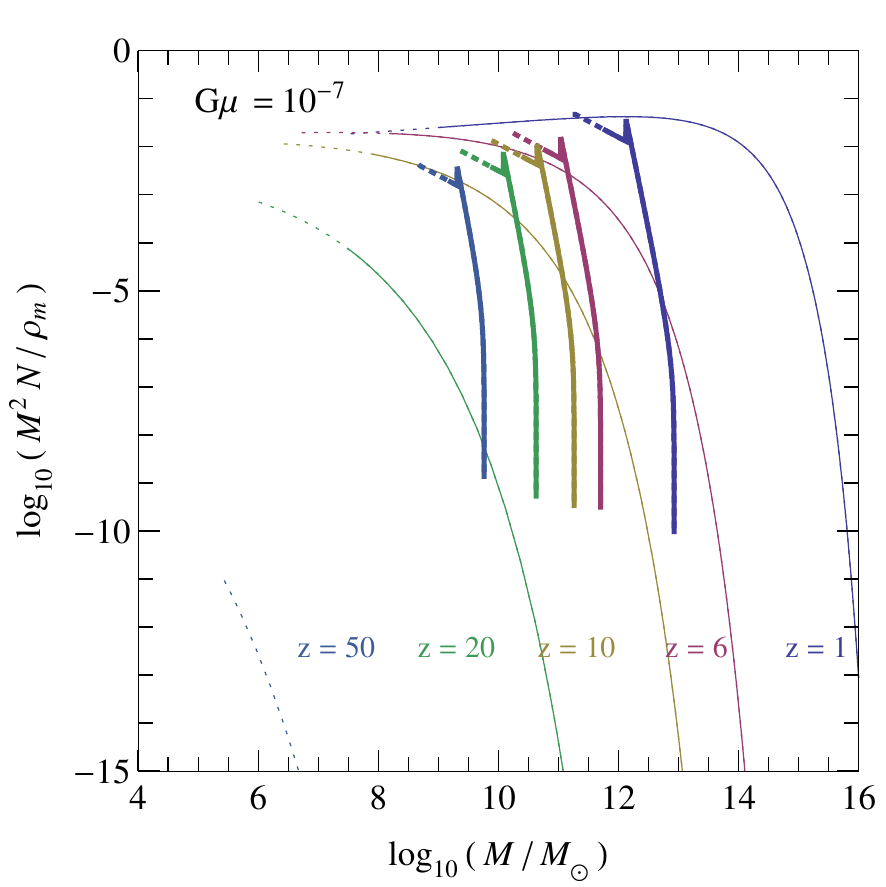}}}\\
      \vspace{-4mm}
    \mbox{\hspace{-.2in}
      \subfigure{\includegraphics[width=3.1in]{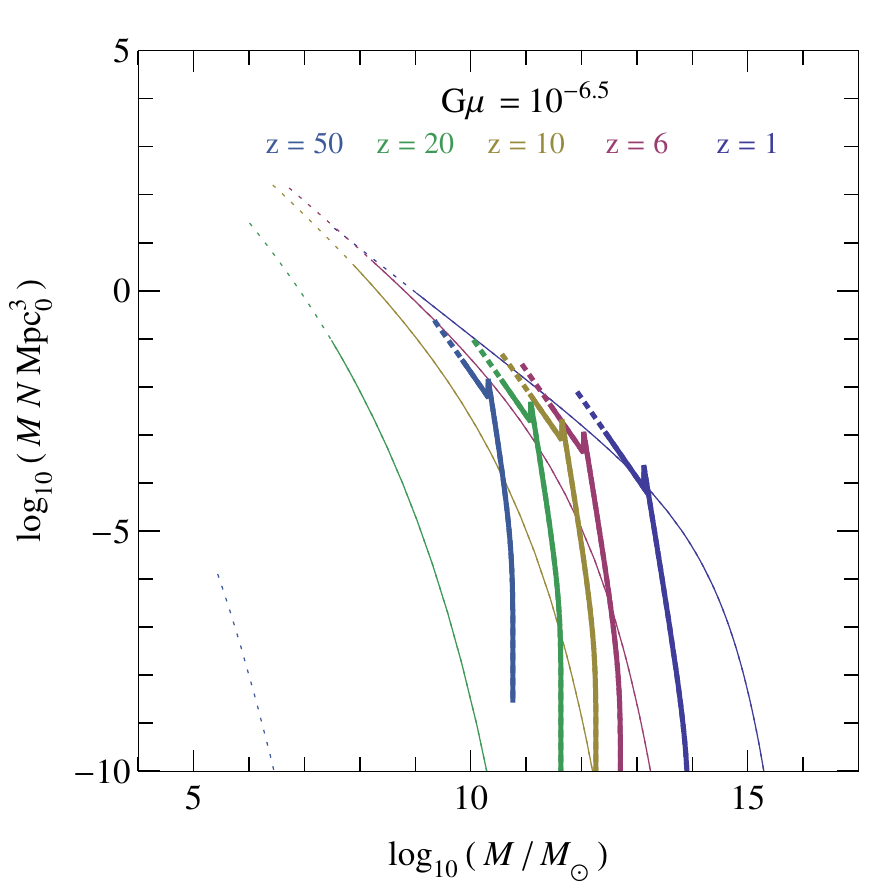}}
      \subfigure{\includegraphics[width=3.1in]{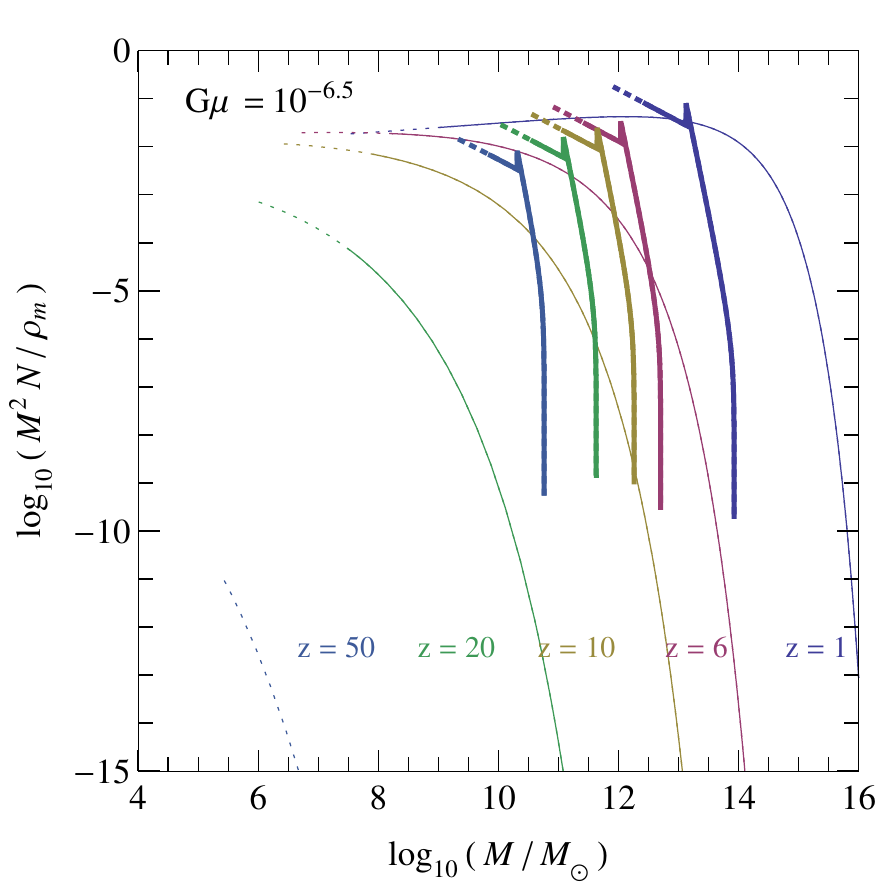}}}\\

   \caption{Number density (left) and mass density (right) in halos with $T_{\rm vir} > 10^4$K (solid) and $T_{\rm vir} > 10^3$K (dotted)  for strings of tension $G\mu = 10^{-7}$ (top) and $G\mu = 10^{-6.5}$ (bottom) .} 
   \label{simple-MdN}
\end{figure}

Numerically, we find
\beq
\label{MN}
M N_{\rm acc}(a,M) \mbox{Mpc}_0^3 &=& 4.9\times 10^{24} \alpha_v\beta\dP
\left(\frac{\alpha \mu_{-8}^2 a^2}{\beta\alpha_v^2 }\right)^{19/3} \left(\frac{M}{M_\odot}\right)^{-7} ,\\
M N_{\rm norm}(a,M) \mbox{Mpc}_0^3 &=& 2.8\times10^8 \dP\frac{\alpha}{\alpha_v} a^2\mu_{-8}^2 
\left(\frac{M}{M_\odot}\right)^{-{5}/{3}},
\eeq
where the unit ${\rm Mpc_0}  = {\rm Mpc}/(1 + z_{\rm eq})$ is the comoving length 
which is a megaparsec today.  We plot these distributions vs. Sheth-Tormen in figures \ref{fig:st75} and \ref{simple-MdN} for three different values of $G\mu$.

In the delta-function approximation for $g$, the largest contribution to the number density of halos is from the smallest loops (since their number density is larger).  This is cut off when the
resulting halos are too small to form stars.  As we consider larger halos, we consider larger loops still in the normal growth regime.  This is cut off when the loops are moving too fast to be in the normal growth regime.  Even larger loops are responsible for accelerated growth halos.
The small discontinuity in the graphs at the transition to this regime is caused by the discontinuity in the Jacobian factor $({\partial m}/{\partial M})$.  The sharp decline of the halo distribution in the accelerated growth regime reflects the steep $M^{-7}$ dependence in (\ref{MN}).

From the figures, there is a clear enhancement of early star formation provided $G\mu \gtrsim 10^{-7}$ if $T_\ast = 10^4$ K, and even for $G\mu \gtrsim 10^{-7.5}$ if $T_\ast = 10^3$ K.  This enhancement occurs for redshifts $z \gtrsim 10-20$, and is limited to a rather narrow range of halo masses. 
Within the delta-function approximation for $g$, there is no early
star formation for $G\mu \leq 10^{-8}\left(\frac{T_\ast}{10^4{\rm K}}\right)$, although in fact, the low momentum tail of $g$ will produce some early stars.

\section{Baryon collapse fraction}\label{sec:collapse}

The fraction of baryons which collapse into stars is proportional to the fraction of matter in star-forming halos, i.e., those of sufficient virial temperature for efficient atomic hydrogen cooling.
Because the process of filament collapse does not significantly affect the virial temperature of the gas, we can neglect this process and simply count the fraction
of matter in filaments of sufficient mass to form stars, i.e., those of mass $M_{\rm fil} \geq M_{\rm fil}^\ast$. The collapse fraction is
\beq
F_{\rm col} &=& \frac{1}{\rho_{\rm m}}\int_{M_{\rm fil}^\ast}^\infty dM_{\rm fil} M_{\rm fil} N_{\rm fil}(M_{\rm fil}),\\
&=& \frac{1}{\rho_{\rm m}}\int_{m_\ast}^{\infty} M_{\rm fil}(m) n(a_{\rm eq},m,v_{\rm eq}) dm dv_{\rm eq},\\
&\approx& 6\pi G \teq^2\int_{m_\ast}^\infty M_{\rm fil}(m)\frac{\dP \mu^{3/2}\alpha^{1/2}}{2 (2 \teq)^{3/2}m^{5/2}}dm,\\
&\approx& \frac{3\pi a G\dP \sqrt{\teq \alpha \mu^3}}{2\sqrt{2}}\times\left\{
\begin{array}{lr}
m_{\rm acc}  \mathlarger{\int_{m_\ast}^\infty  \frac{dm}{m^{5/2}}}&\qquad m_{\rm acc} \leq m_\ast\\
\mathlarger{\int_{m_\ast}^{m_{\rm acc}}\frac{dm}{m^{3/2}} +  m_{\rm acc} \int_{m_{\rm acc}}^\infty  \frac{dm}{m^{5/2}}}&m_{\rm acc} \geq m_\ast
\end{array}
\right.,\label{eq:Fcol-norm-acc}
\eeq
where $m_\ast$, the smallest loop mass responsible for star forming filaments, is determined from $T_{\rm vir} \geq T_\ast$ and (\ref{eq:fil-T-vir}) to be
\beq\label{eq:m-star}
m_\ast = \frac{18\Meq T_\ast^2\alpha_v^2}{G\mu\alpha m_p^2},
\eeq
and the smallest loop mass undergoing accelerated growth is given by
\beq
m_{\rm acc} = \left(\frac{2\alpha a^2 (G\mu)^5}{\beta^4\alpha_v^2}\right)^{1/3} \Meq.
\eeq
The filament mass $M_{\rm fil}(m)$ from (\ref{eq:filament-mass}) then takes the form
\beq
M_{\rm fil}(m) = \left\{ 
\begin{array}{ll}
a m & m \leq m_{\rm acc}\\
a m_{\rm acc}\quad& m \geq m_{\rm acc}.
\end{array}
\right.
\eeq

It will be convenient to define the time before which no star-forming filaments are in the normal-growth regime by $m_\ast = m_{\rm acc}$, yielding
\beq
a_{\rm acc} = \frac{54 T_\ast^3 \alpha_v^4\beta^2}{(G\mu)^4m_p^3\alpha^2} = \frac{1.4\times 10^5}{\mu_{-8}^4}\left(\frac{T_\ast}{10^4 {\rm K}}\right)^3.
\eeq
Then we can write
\beq
F_{\rm col} = \left\{
\begin{array}{ll}
\dfrac{3\pi G a\dP\sqrt{\teq \alpha\mu^3}}{\sqrt{2}}\dfrac{m_{\rm acc}}{3m_\ast^{3/2}}\quad&a < a_{\rm acc}  \vspace{3mm}\\
\dfrac{3\pi G a\dP\sqrt{\teq \alpha\mu^3}}{\sqrt{2}}\left(\dfrac{1}{\sqrt{m_\ast}}- \dfrac{2}{3\sqrt{m_{\rm acc}}}\right)\quad&a \geq a_{\rm acc} .\label{eq:Fcol-late}
\end{array}
\right.
\eeq

It is here where we examine the consequences of using the delta-function approximation for the loop production function $g$.  In particular, we have found that the large typical loop velocity does limit early star formation, and so we should verify that rare slow loops are not more important.   Notice that the low-velocity tail of the distribution will rapidly diminish the value of $a_{\rm acc} \propto \alpha_v^4$.  This means the normal-growth dominated $(a > a_{\rm acc})$ part of (\ref{eq:Fcol-late}) is relevant, and so
\beq
F_{\rm col} \propto \dP\frac{\alpha}{\alpha_v}.
\eeq
Simulations reveal a loop production function $g(\alpha,\alpha_v)$ consisting of several features, including a scaling peak and a broad plateau (see figures 5 \& 7 in \cite{BOS}).   This distribution becomes rather localized after multiplication by $\alpha$ and division by $\alpha_v$, since all features except the scaling peak become suppressed.  The numerical results are well fit by a single localized peak for the distribution
\beq
\frac{\alpha}{\alpha_v}g(\alpha,\alpha_v) \approx 7 \frac{0.05}{0.3}\delta(\alpha - 0.05)\delta(\alpha_v - 0.3), 
\eeq
and hence, for the purpose of calculating $F_{\rm col}$,
\beq
g(\alpha,\alpha_v) \approx 7\,\delta(\alpha - 0.05)\delta(\alpha_v - 0.3).
\eeq
For the potentially observable range of $G\mu$ and $T_\ast$ we can write
\beq
F_{\rm col} \approx \frac{\pi}{2}a\dP\frac{\alpha m_p}{\alpha_vT_\ast}(G\mu)^2 \approx 6.5\times 10^{-4}\left(\frac{10^4{\rm K}}{T_\ast}\right)\frac{\mu_{-8}^2}{1+z} .
\eeq

\section{Ionization history}\label{sec:ionization}

\begin{figure}[htbp] 
   \centering
   \mbox{\hspace{-.2in}
   	\subfigure{\includegraphics[width=3.7in]{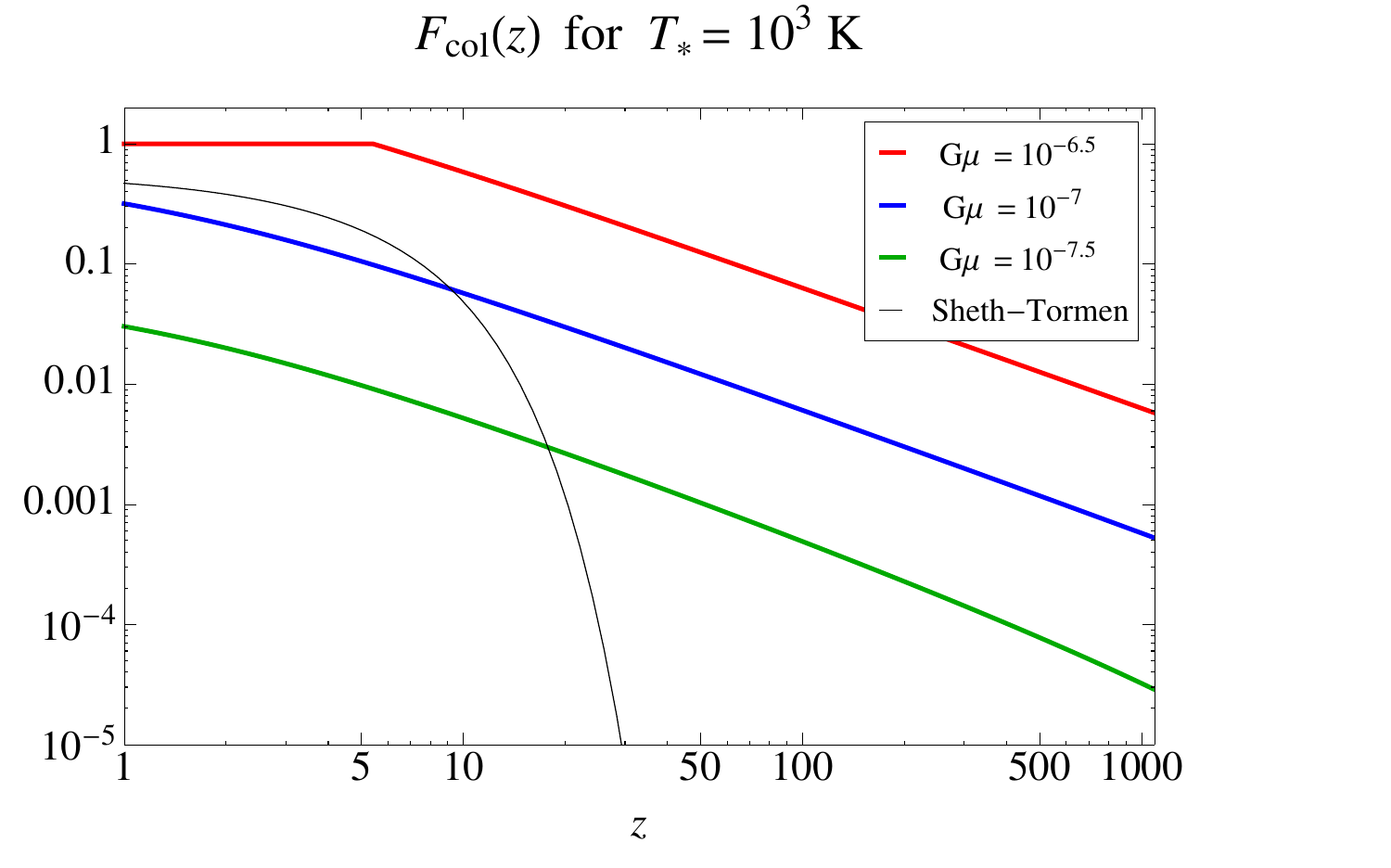}}
	{\hspace{-.6in}}
	\subfigure{\includegraphics[width=3.7in]{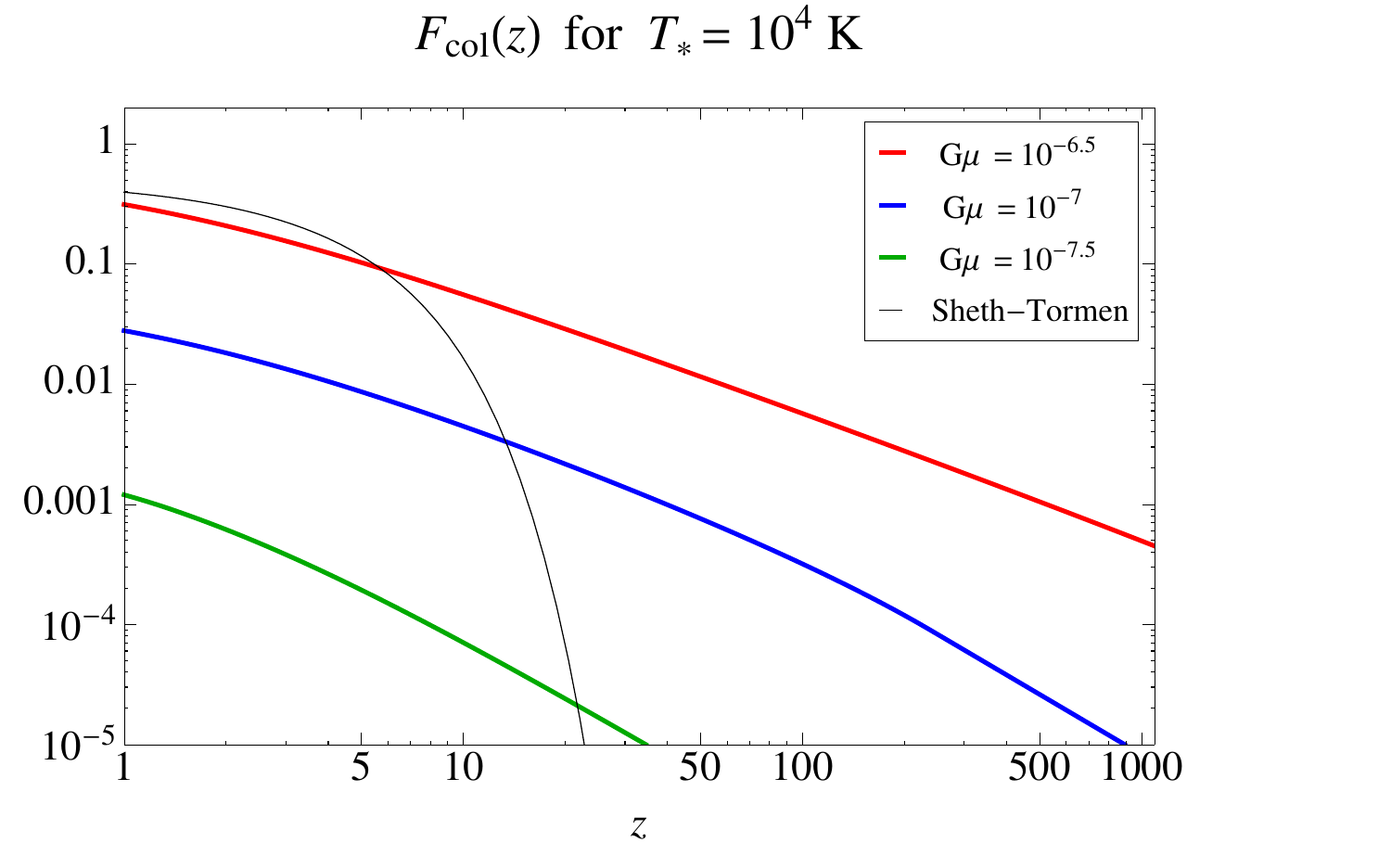}}
	}\\ 
   \mbox{\hspace{-.2in}
	\subfigure{\includegraphics[width=3.7in]{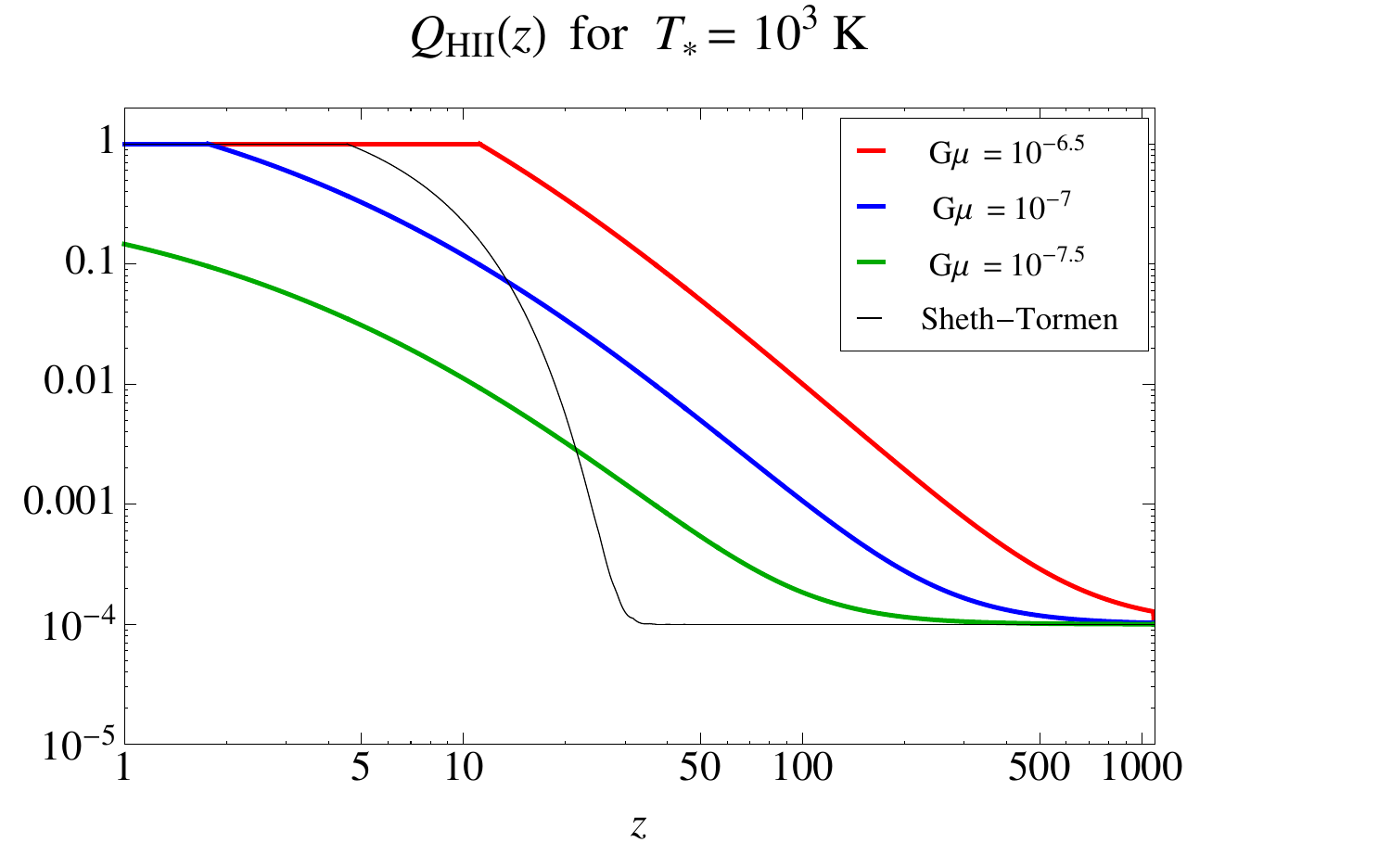}}
	{\hspace{-.6in}}
	\subfigure{\includegraphics[width=3.7in]{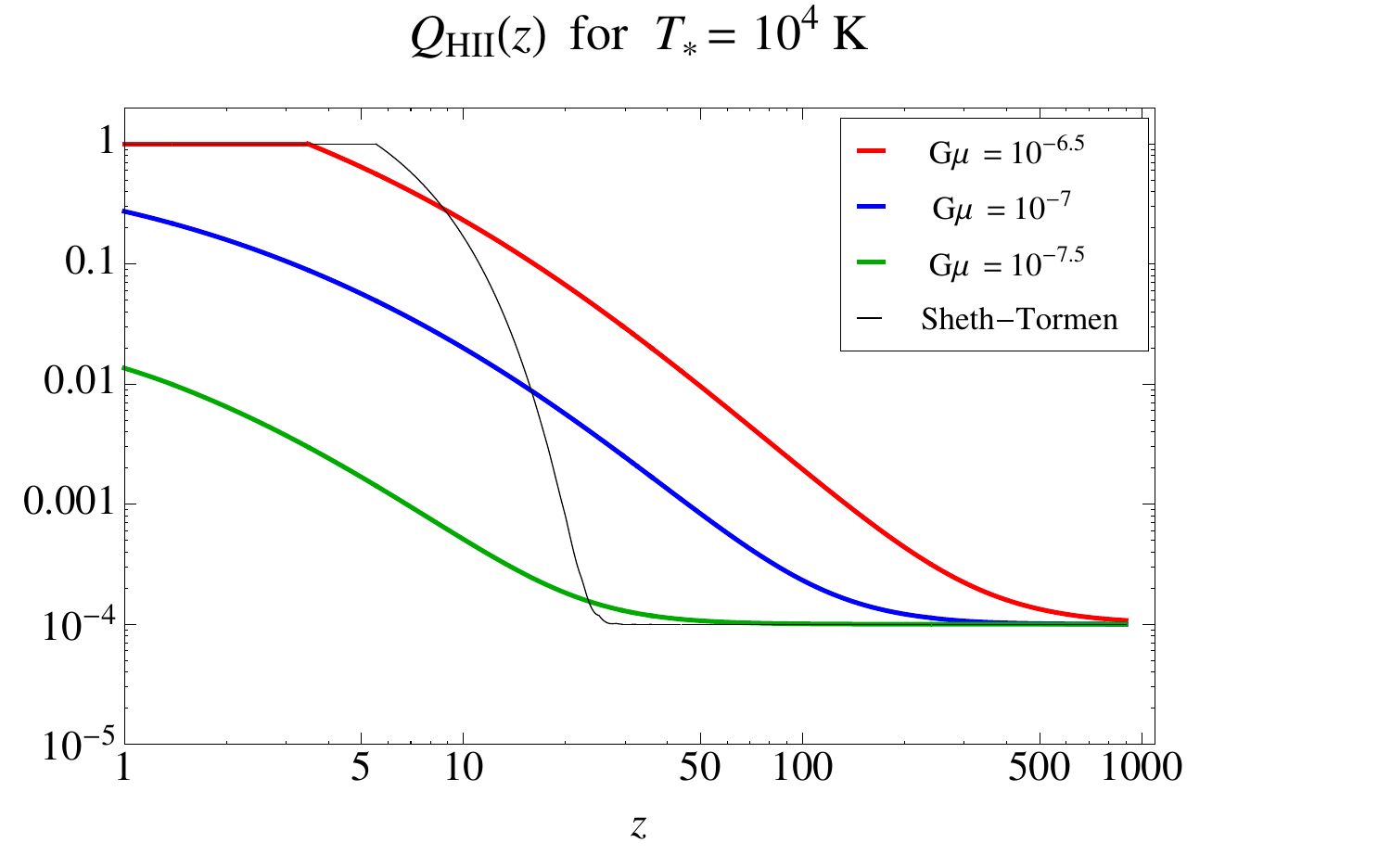}}
   	     }
   \caption{Fraction of matter in loop-seeded star-forming halos (top) and fraction of volume which has thereby been re-ionized (bottom) as a function of redshift.   We include a residual floor of $10^{-4}$ in $Q_{\rm HII}$.  The Sheth-Tormen scenario is shown for comparison (thin black  line).} 
   \label{fig:Fcollapse}
\end{figure}

The fraction of volume which has been re-ionized is related to the
baryon collapse fraction $F_{\rm col}$ via \cite{BL01, WL03}
\beq 
Q_{\rm HII}(a) = \int_0^a\frac{N_{\rm ion}}{0.76}\frac{dF_{\rm col}}{da'}e^{F(a',a)}da', 
\eeq 
where the number of ionizing photons
per baryon in virialized halos $N_{\rm ion}$ depends on the mass
distribution of stars.  For the initial mass function of stars
observed in nearby galaxies\footnote{The value of $N_{\rm ion}$
increases by $1.5$ orders of magnitude if the stars are all more
massive than $100M_\odot$ \cite{BKL}.},
$N_{\rm ion} = 40 (f_\star/0.1) (f_{\rm esc}/0.1)$, where $f_\star$ is the
efficiency of converting baryons in virialized halos into stars and
$f_{\rm esc}$ is the fraction of ionizing photons which escape the
halo.  Here \beq F(a',a) = -0.26\frac{C}{10}\left[\left(\frac{1+z_{\rm
eq}}{a'}\right)^{3/2} - \left(\frac{1+z_{\rm
eq}}{a}\right)^{3/2}\right], \eeq where $C = \langle n_{\rm H}^2\rangle/\langle n_{\rm H}\rangle^2$ is the clumpiness.  We can
evaluate $Q_{\rm HII}(z)$ in closed form using the incomplete Gamma
function, although the result is rather long, so we will not present
it explicitly here.  The clumping factor $C$ is expected to approach unity at high redshifts \cite{Trac}; hence we set $C=1$. 
We plot $F_{\rm col}(z)$ and $Q_{\rm HII}(z)$ for various scenarios in figure \ref{fig:Fcollapse} below using\footnote{These values of $N_{\rm ion}$ are chosen for consistency with the WMAP optical depth; see figure \ref{fig:nion} below.}
$N_{\rm ion} = 10$ for $T_*=10^4 {\rm K}$ and $N_{\rm ion} = 5$ for $T_*=10^3 {\rm K}$.

We can find the optical depth by integrating the reionization fraction out to redshift $z$ via 
\beq
\tau(z) = \int_0^z n_{\rm H}(z')Q_{\rm HII}(z')\sigma_{\rm T}{\frac{dz'}{(1+z')H(z')}},
\eeq
where $n_{\rm H}(z)$ is the physical number density of hydrogen, equal to $2.7\times 10^{-7}\mbox{cm}^{-3}$ today, and $\sigma_{\rm T} = 6.652\times 10^{-25}\mbox{cm}^2$ is the Thomson cross-section.  The Hubble rate obeys $H(z) = H_0\sqrt{\Omega_\Lambda + \Omega_m(1 + z)^3}$, with $\Omega_\Lambda = 0.728$ and $\Omega_m = 1-\Omega_\Lambda$.

\begin{figure}[htbp] 
   \centering
   \mbox{\hspace{-.1in}\subfigure{
   \includegraphics[width=3.5in]{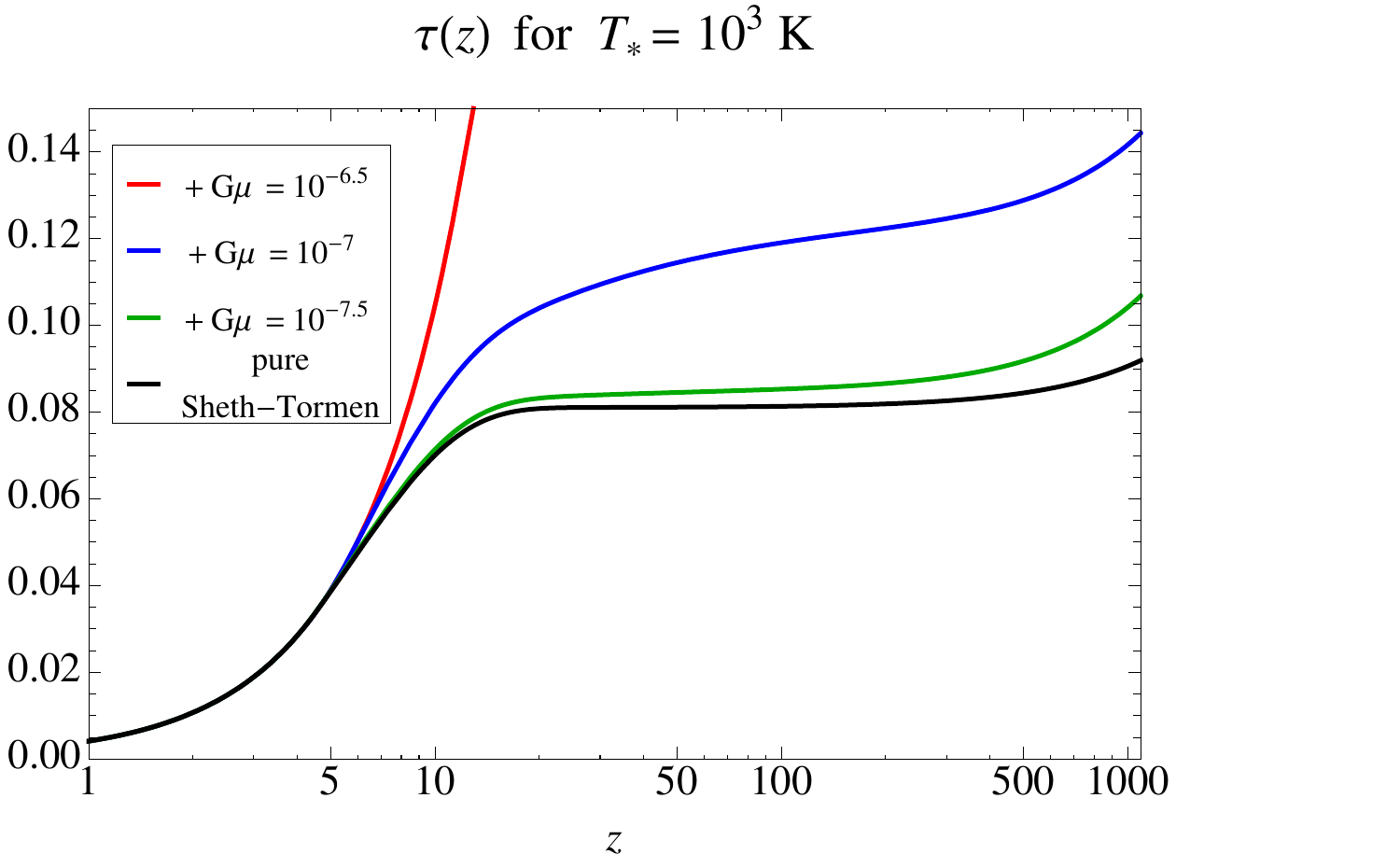} 
   }{\hspace{-.4in}}\subfigure{
     \includegraphics[width=3.5in]{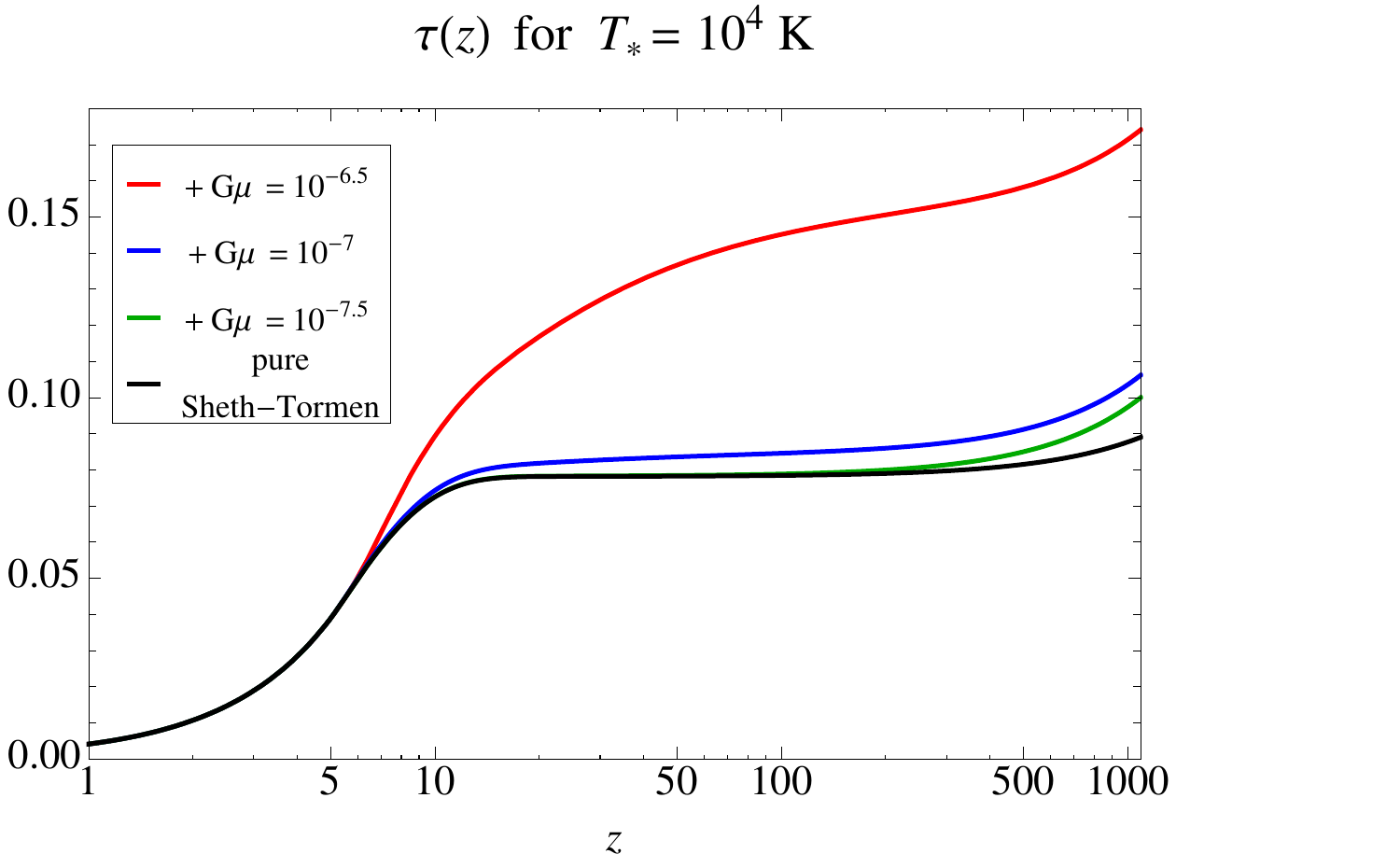}
    }}
   \caption{Optical depth from Sheth-Tormen plus cosmic strings for two possible values for $T_\ast$.} 
   \label{fig:tau}
\end{figure}

\begin{figure}[htbp] 
   \centering
   \mbox{\hspace{-.1in}\subfigure{
   \includegraphics[width=3.5in]{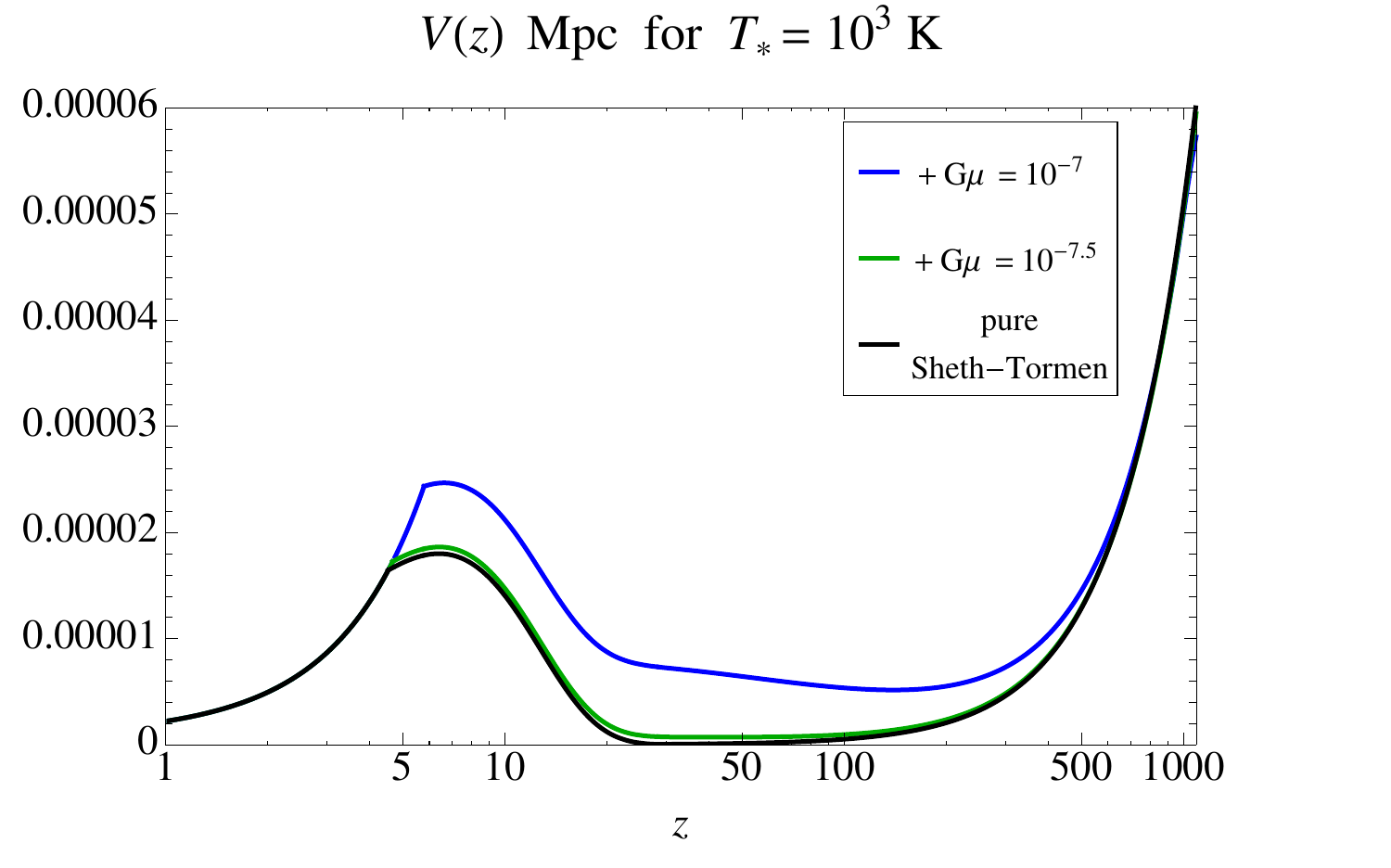} 
   }{\hspace{-.4in}}\subfigure{
     \includegraphics[width=3.5in]{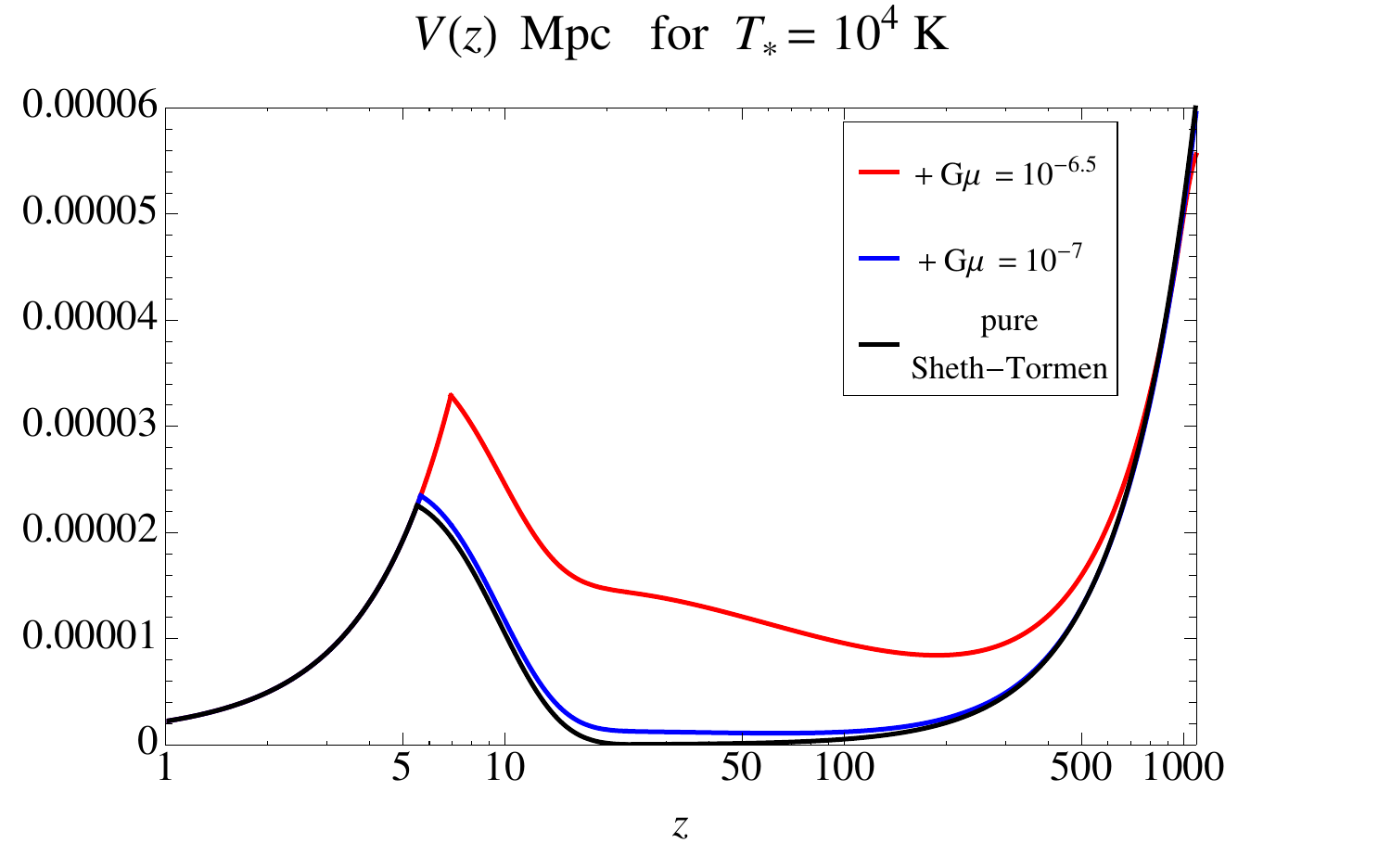}
    }}
   \caption{Visibility function from Sheth-Tormen plus cosmic strings for two possible values for $T_\ast$. } 
   \label{fig:visibility}
\end{figure}

The visibility function $V(z)$ is defined through the optical depth $\tau$ via,
\beq
V(z) = e^{-\tau(z)}{H(z)\frac{\partial\tau}{\partial z}}.
\eeq
We plot $\tau(z)$ and $V(z)$ in figures \ref{fig:tau} \& \ref{fig:visibility}.

Using the WMAP+BAO+$H_0$ \cite{Komatsu:2010fb} measured value of the reionization optical depth $\tau = 0.087 \pm 0.014$, we can find the excluded 
region\footnote{The parameters indicated in figure \ref{fig:nion} are excluded only within the framework of our simplified model, which assumes that $N_{\rm ion}$, $T_\ast$, and $C$ are constants independent of the redshift.  In particular, larger values of $N_{\rm ion}$ may be allowed at smaller redshifts, where the clumpiness $C$ is likely to be $>1$.} of $N_{\rm ion}$-$G\mu$ parameter space, shown in figure \ref{fig:nion} below.

Using the functional form of $Q_{\rm HII}(z)$, we examine the effects of early reionization due to strings on the CMB temperature and $E$-mode polarization power spectra in figures \ref{fig:CTT} \& \ref{fig:CEE}, as well as on the TE
cross-correlation in figure \ref{fig:CTE}.  In these figures we plot both the pure Sheth-Tormen case as well as Sheth-Tormen plus cosmic strings for the specified values of $G\mu$ and $T_*$.  
(Note that here we have accounted only for the effect of strings on reionization, disregarding their direct gravitational influence on CMB spectra, e.g., through the Gott-Kaiser-Stebbins effect.)
We additionally plot the fractional difference in power spectra induced by cosmic strings. These figures were obtained using a modification of the CAMB\footnote{ http://camb.info/} software package \cite{Lewis:1999bs}.  Because Planck
is expected to be near cosmic variance limited \cite{:2006uk}, we have shaded the regions of these figures below the cosmic variance detection threshold for a single $C_l$.  (Features consisting of several consecutive $C_l$s can be estimated more accurately.)

\begin{figure}[htbp] 
   \centering
   \mbox{\hspace{-.2in}\subfigure{
   \includegraphics[width=3.0in]{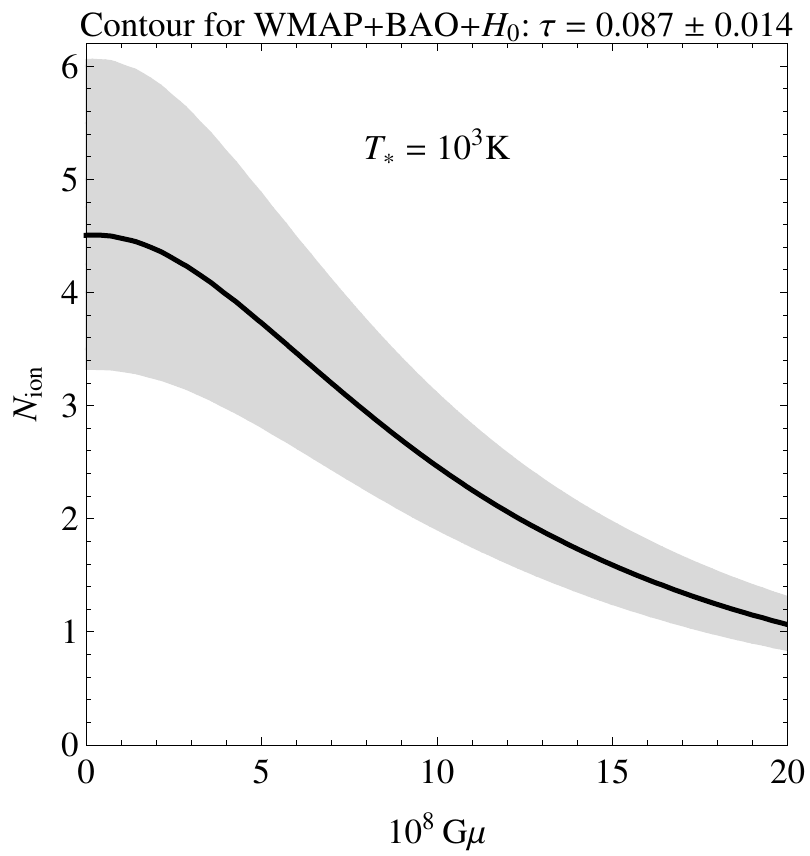} 
   }{\hspace{-0in}}\subfigure{
     \includegraphics[width=3.05in]{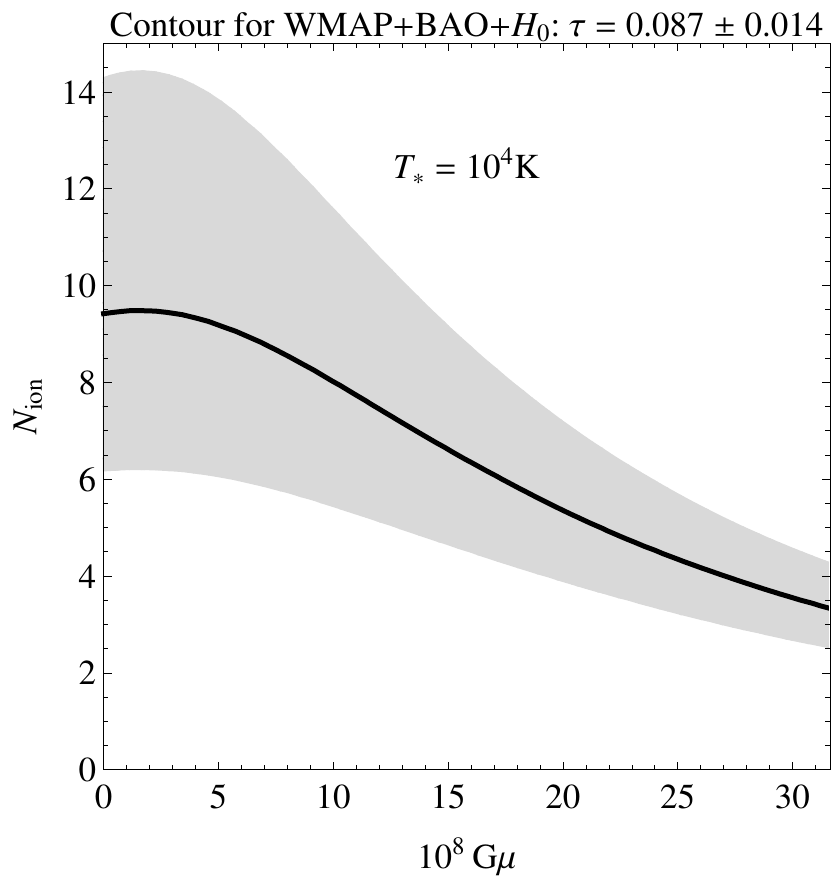}
    }}
   \caption{Predicted optical depth vs. $N_{\rm ion}$ and $G\mu$.  Parameter space shaded with gray is compatible with the 68\% CL WMAP+BAO+$H_0$ \cite{Komatsu:2010fb} reionization optical depth.  Note that our value for optical depth is defined by integrating from $z=0$ to $z_{\rm decoupling}$.  The pure Sheth-Tormen scenario is equivalent to $G\mu = 0$. } 
   \label{fig:nion}
\end{figure}

\begin{figure}[htbp] 
   \centering
   \mbox{\hspace{-.0in}
	\subfigure{\includegraphics[height=2.2in]{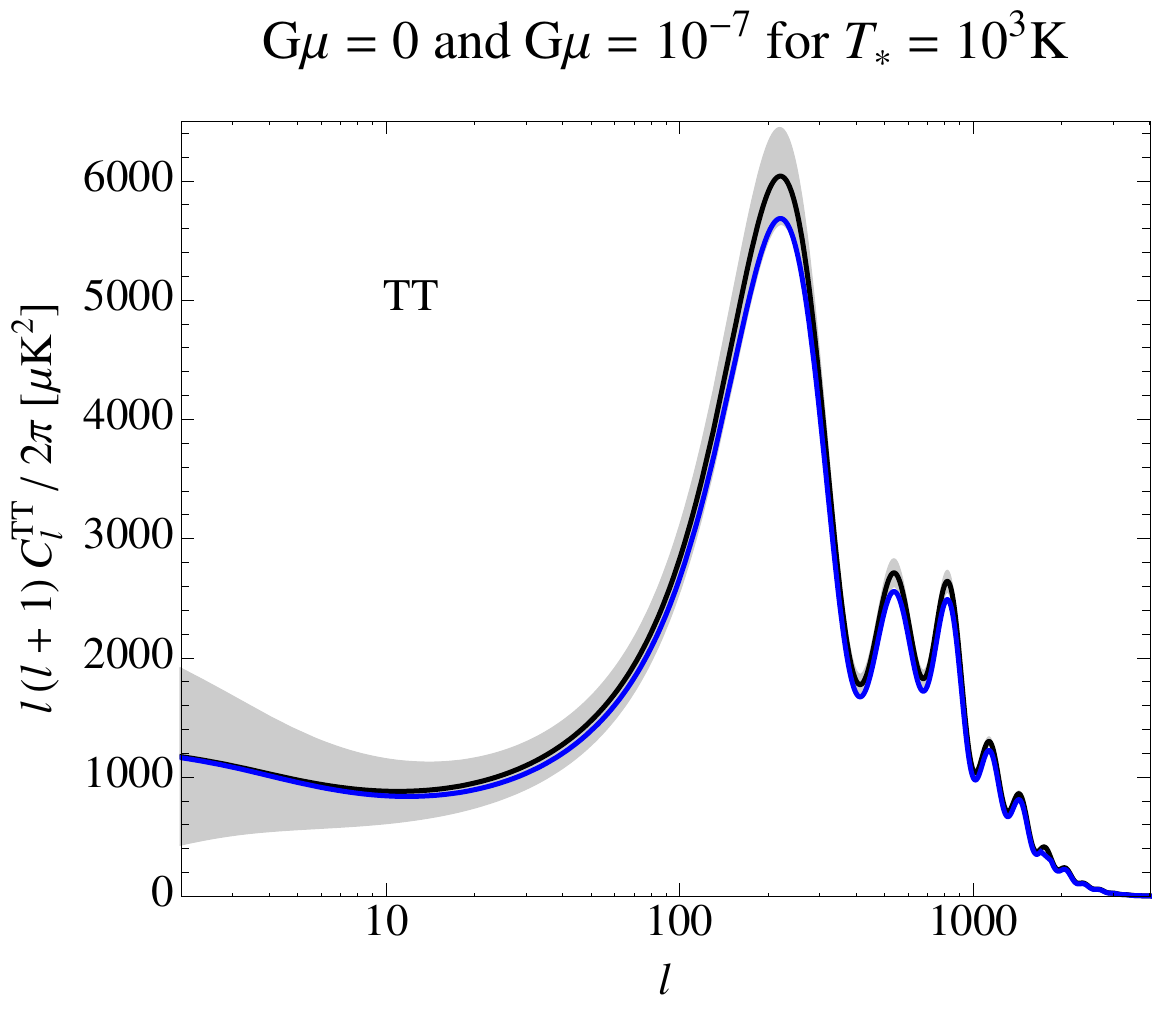}}
	{\hspace{-.0in}}
	\subfigure{\includegraphics[height=2.2in]{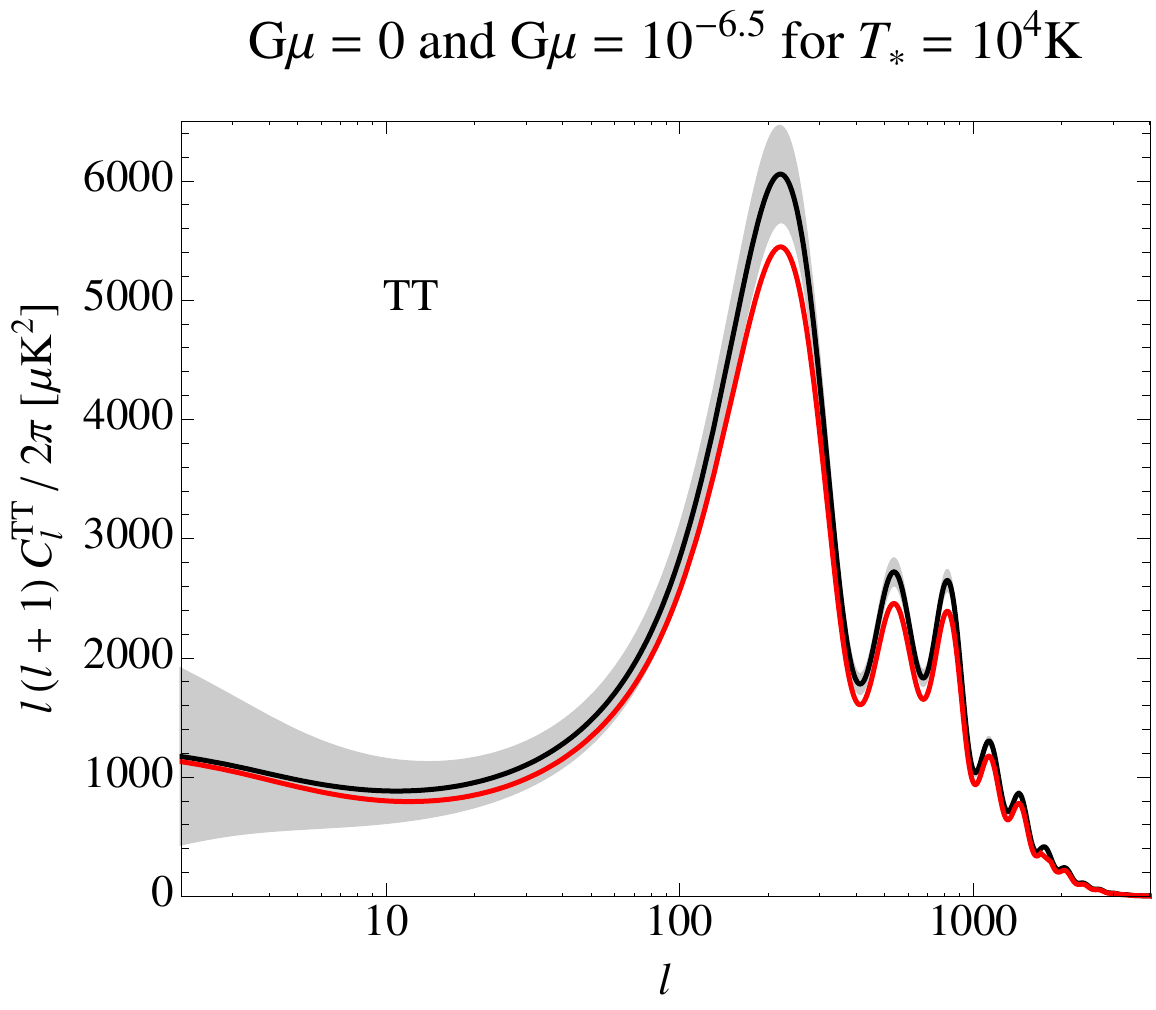}}
   	     }\\
	        \mbox{\hspace{.1in}
	\subfigure{\includegraphics[height=2.2in]{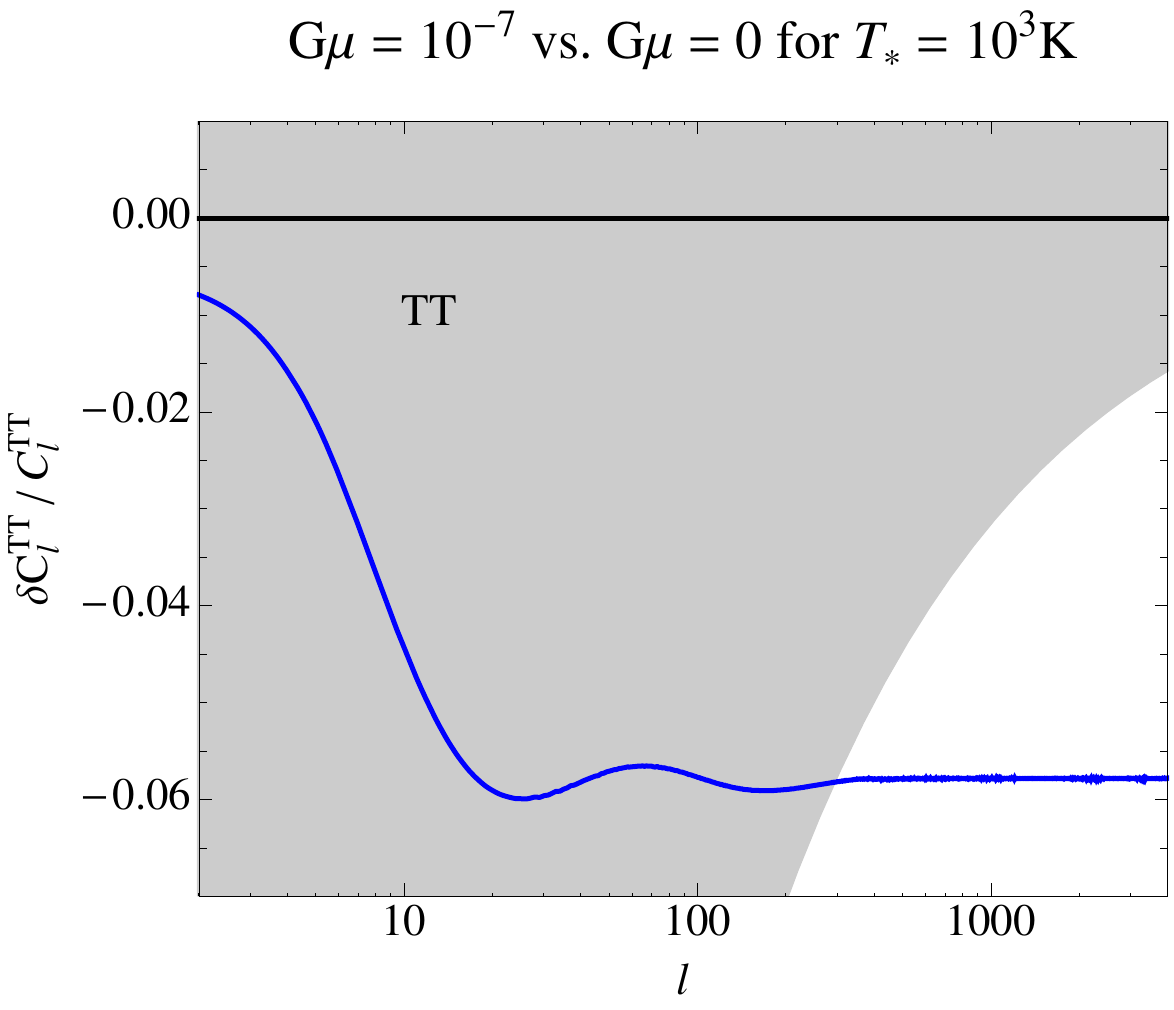}}
	{\hspace{.2in}}
	\subfigure{\includegraphics[height=2.2in]{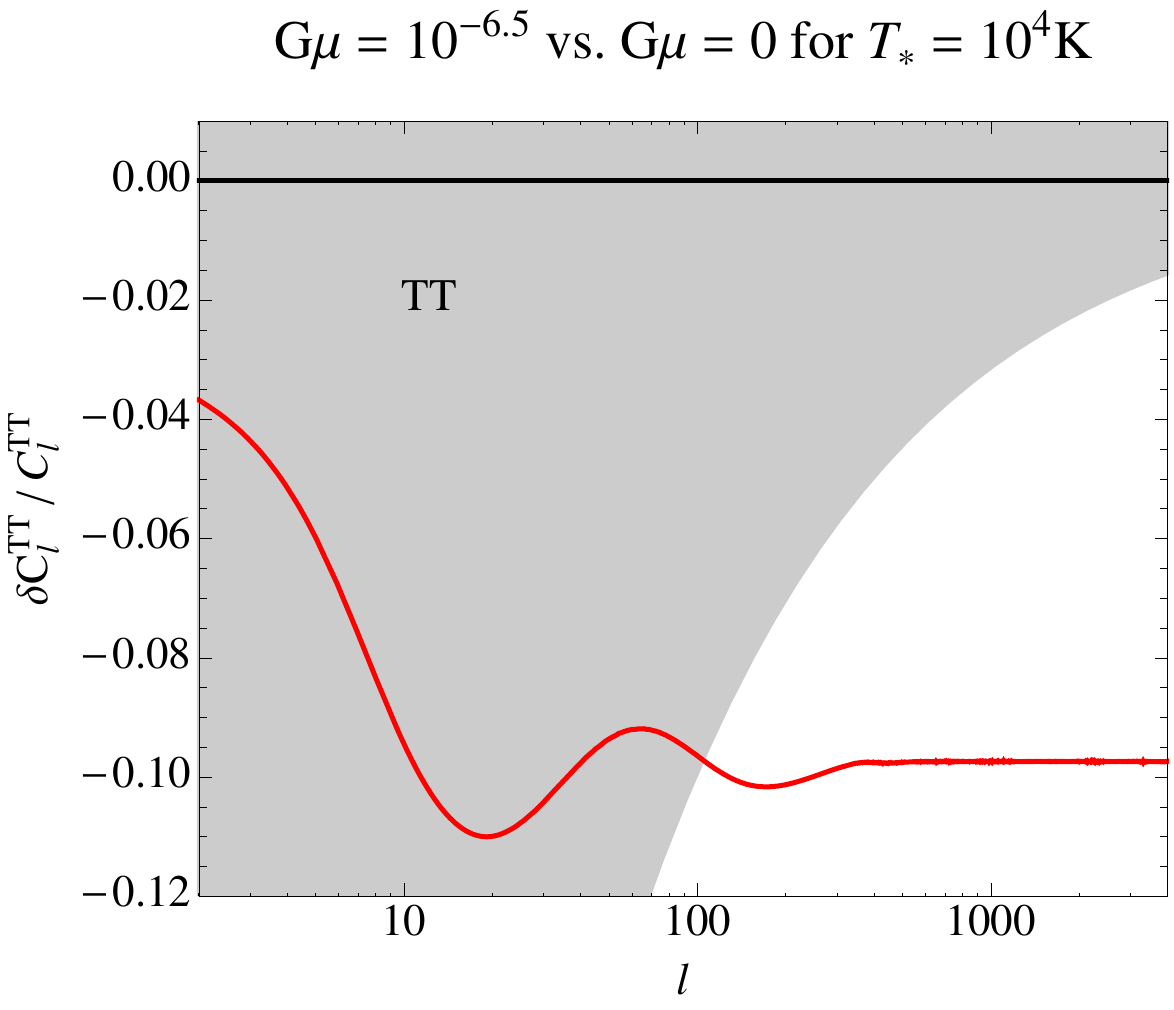}}
   	     }
   \caption{CMB temperature angular power spectra with and without cosmic strings (top) and their fractional difference (bottom).  We assume $T_\ast = 10^3$K (left) and $T_\ast = 10^4$K (right).  The shaded region indicates the single $C_l$ cosmic variance for the $G\mu = 0$ case.} 
   \label{fig:CTT}
\end{figure}

\begin{figure}[htbp] 
   \centering
   \mbox{\hspace{-.0in}
	\subfigure{\includegraphics[height=2.2in]{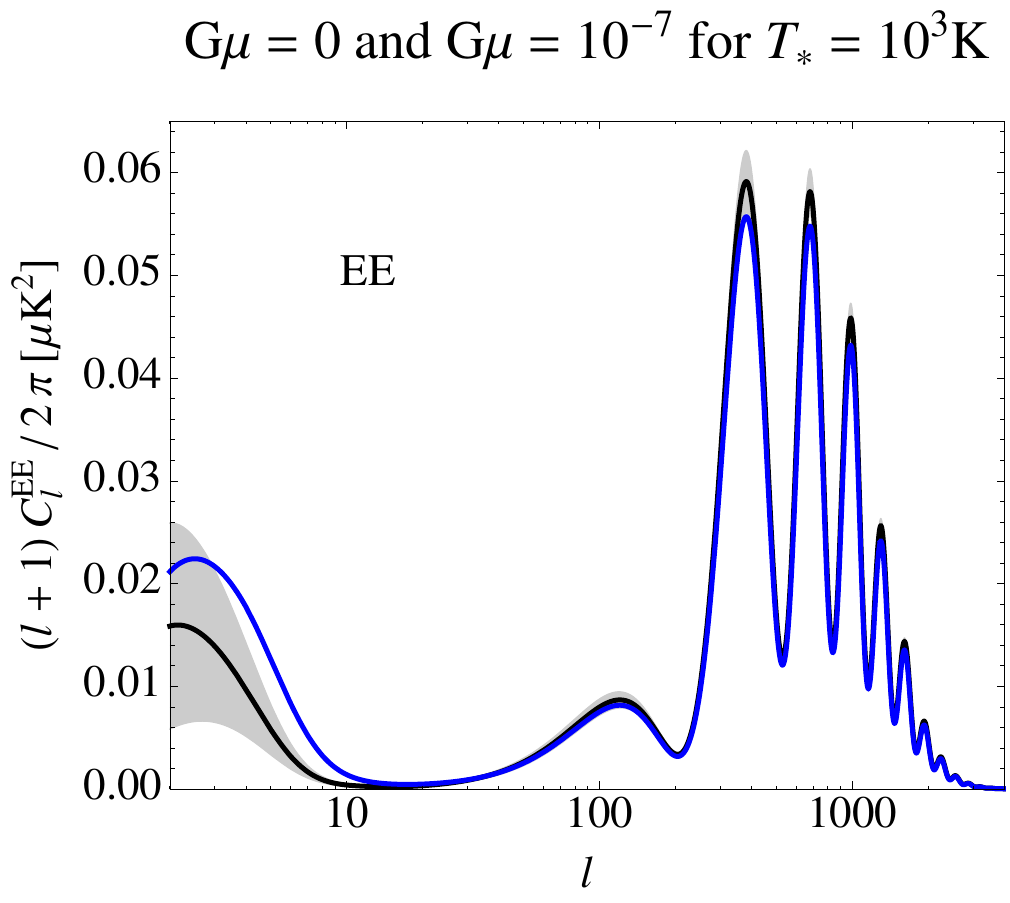}}
	{\hspace{-.0in}}
	\subfigure{\includegraphics[height=2.2in]{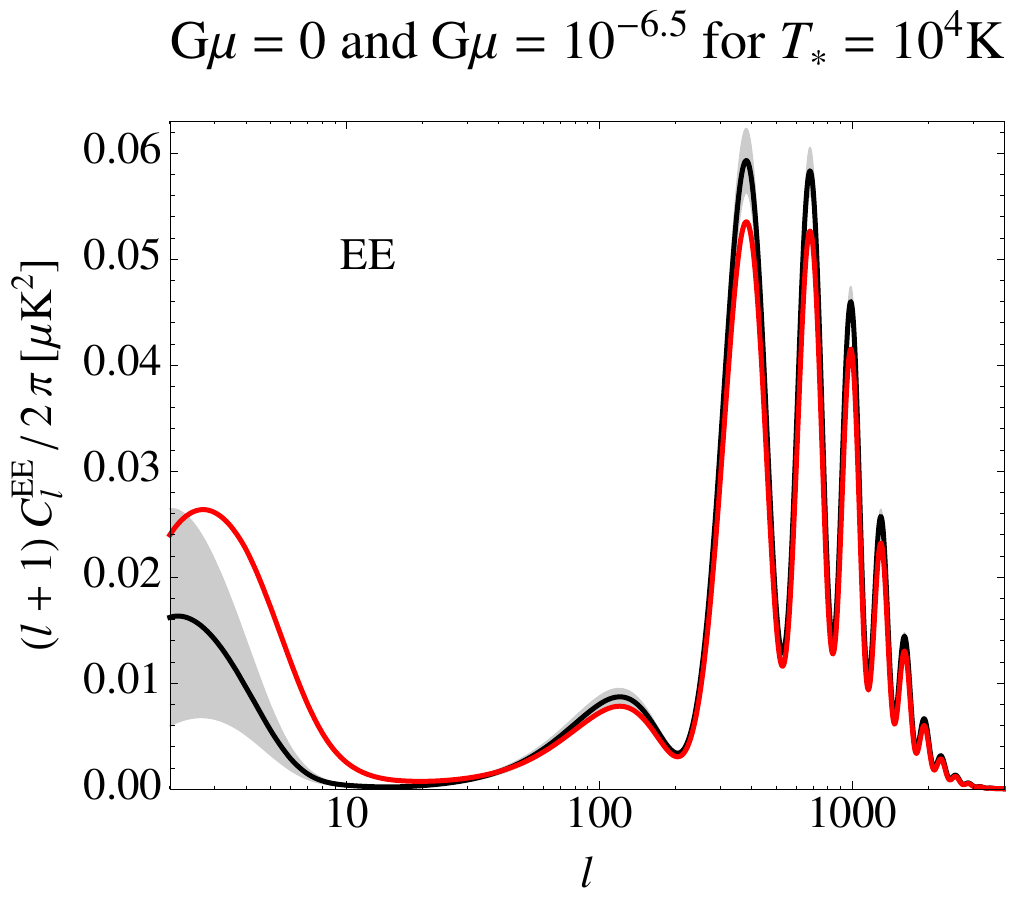}}
   	     }\\
	        \mbox{\hspace{.1in}
	\subfigure{\includegraphics[height=2.2in]{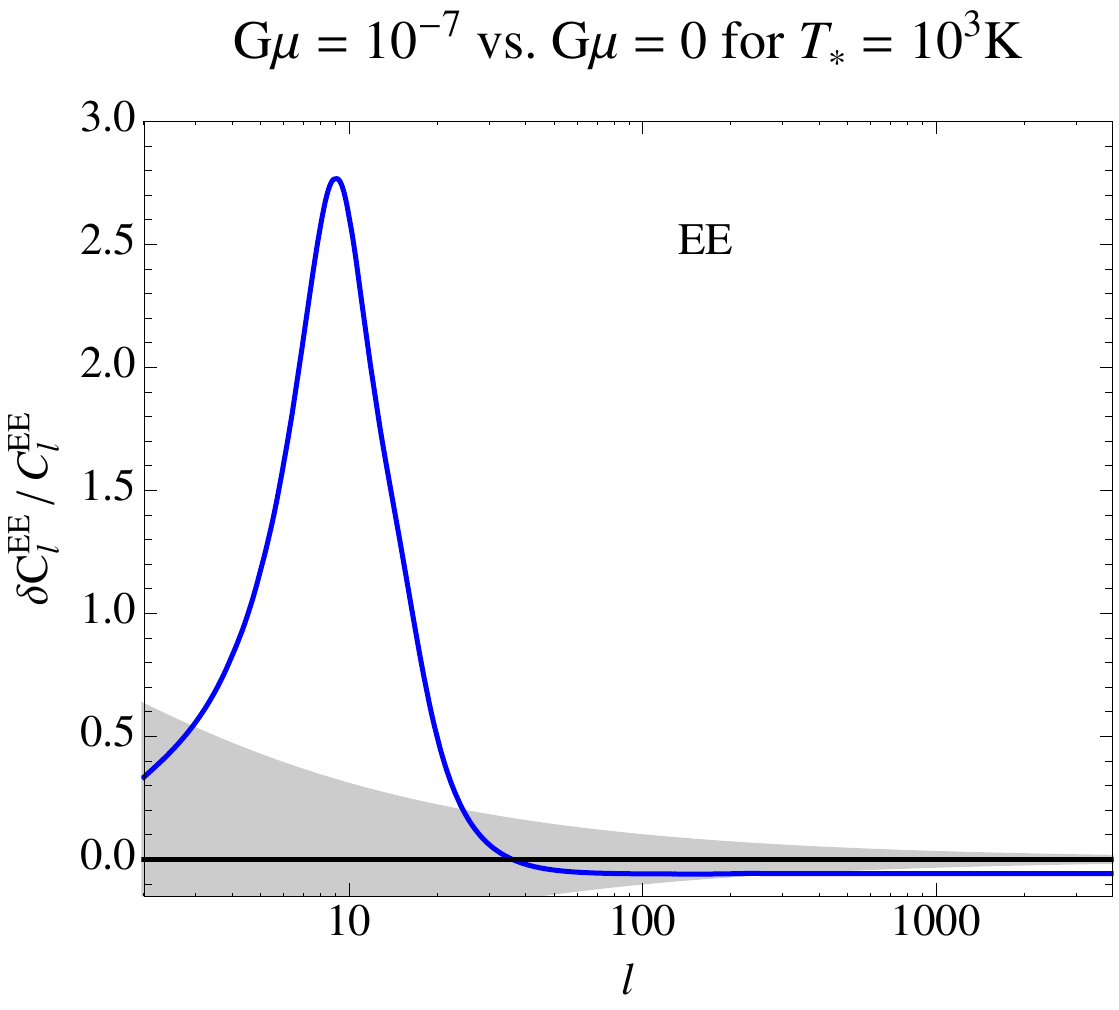}}
	{\hspace{.2in}}
	\subfigure{\includegraphics[height=2.2in]{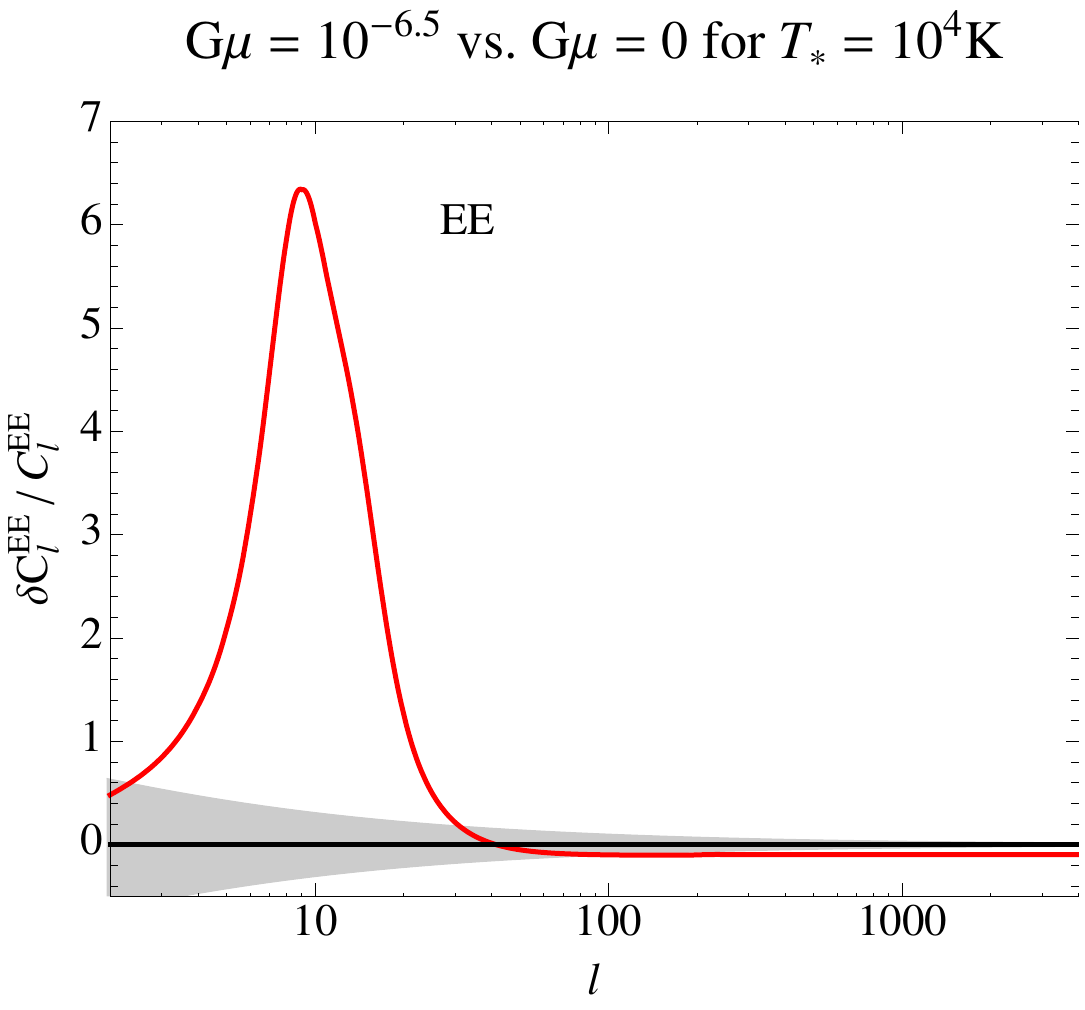}}
   	     }
   \caption{CMB $E$-mode polarization angular power spectra with and without cosmic strings (top) and their fractional difference (bottom).  We assume $T_\ast = 10^3$K (left) and $T_\ast = 10^4$K (right).  The shaded region indicates the single $C_l$ cosmic variance for the $G\mu = 0$ case.} 
   \label{fig:CEE}
\end{figure}

\begin{figure}[htbp] 
   \centering
   \mbox{\hspace{-.0in}
	\subfigure{\includegraphics[height=2.2in]{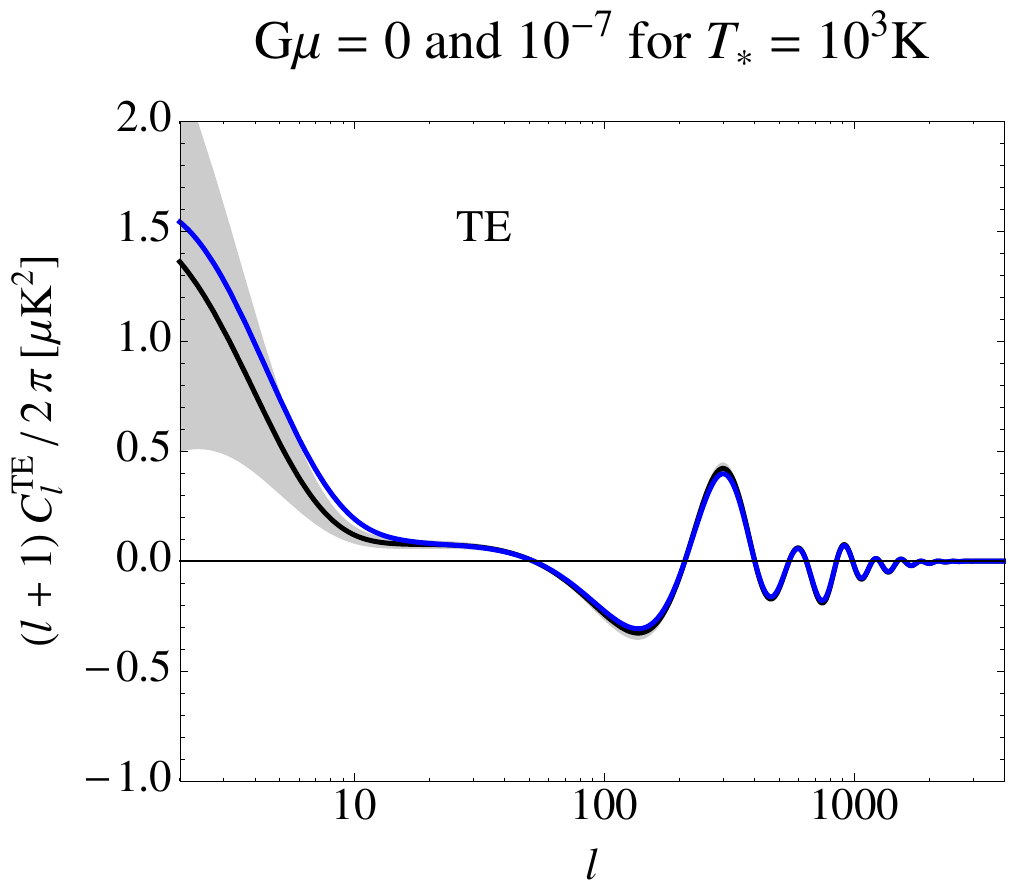}}
	{\hspace{-.0in}}
	\subfigure{\includegraphics[height=2.2in]{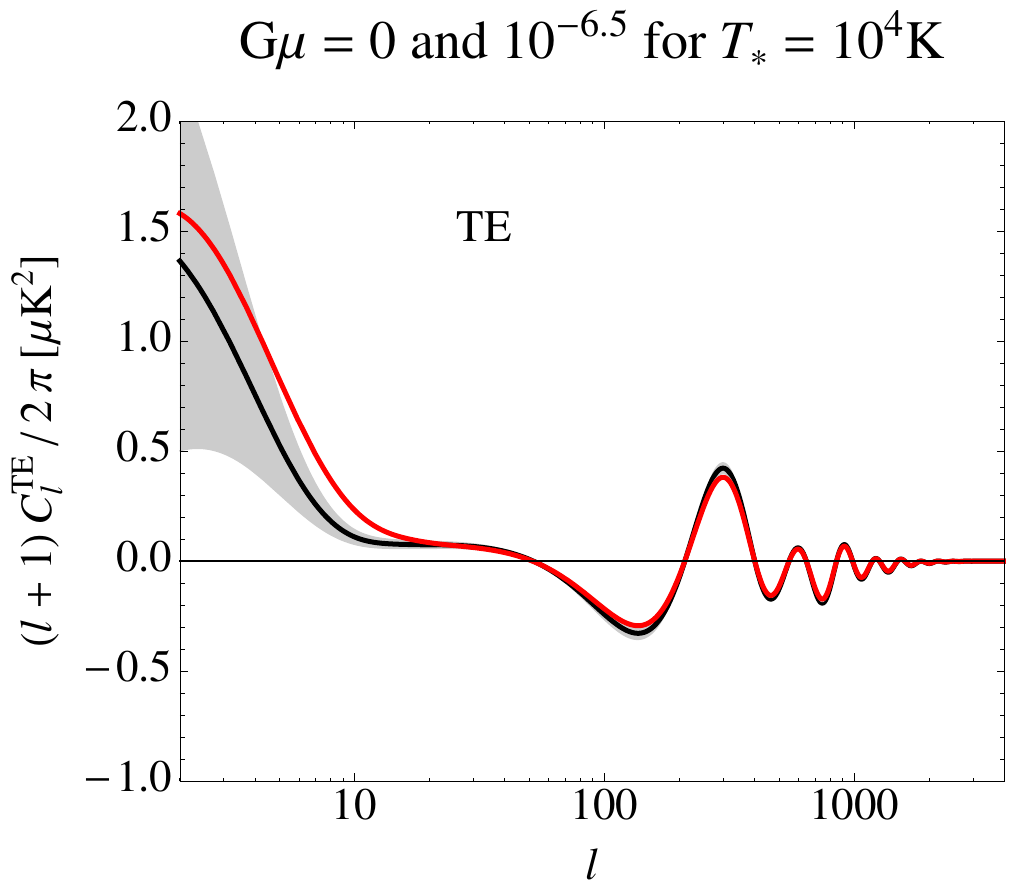}}
   	     }\\
	        \mbox{\hspace{.1in}
	\subfigure{\includegraphics[height=2.2in]{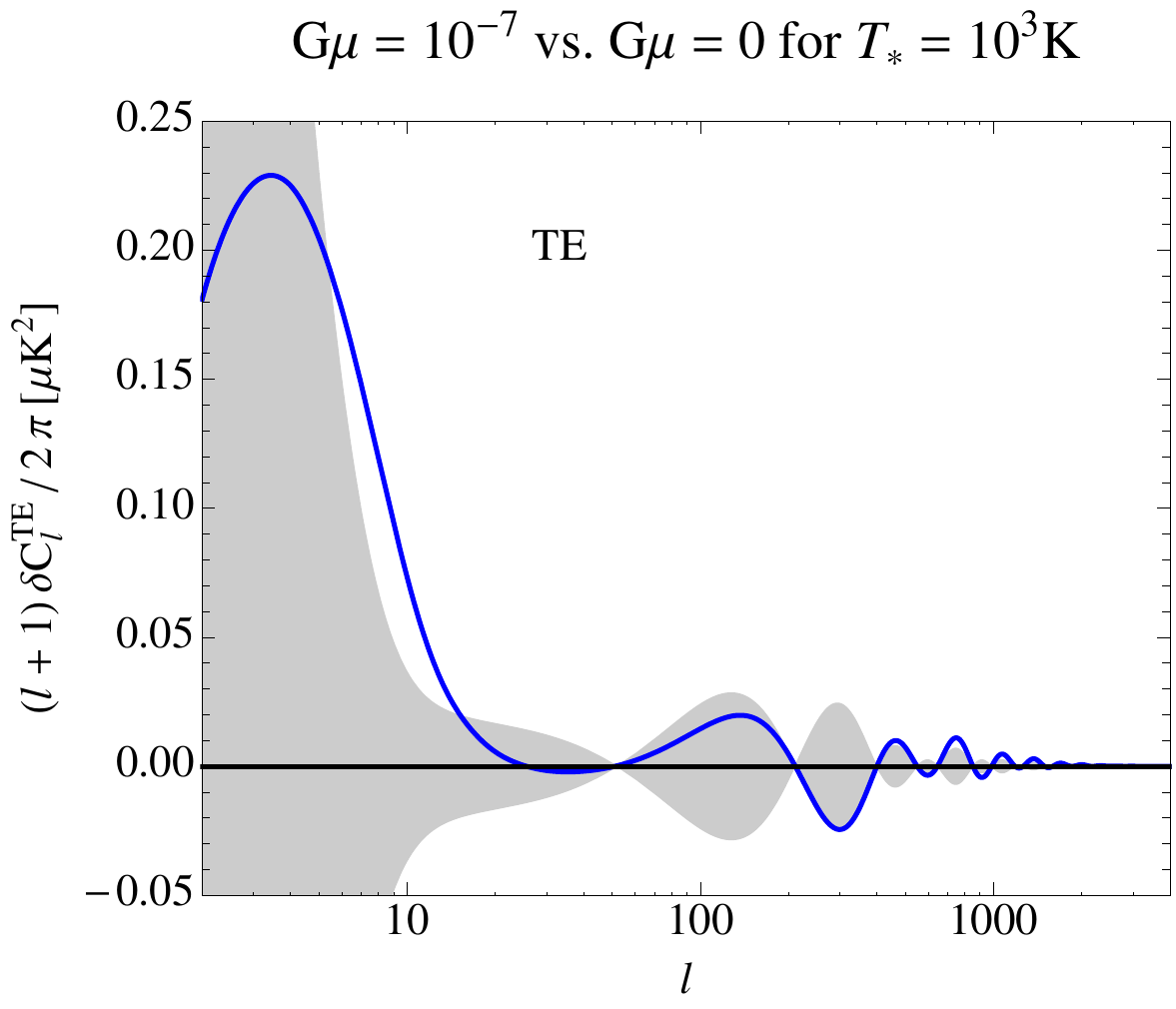}}
	{\hspace{.2in}}
	\subfigure{\includegraphics[height=2.2in]{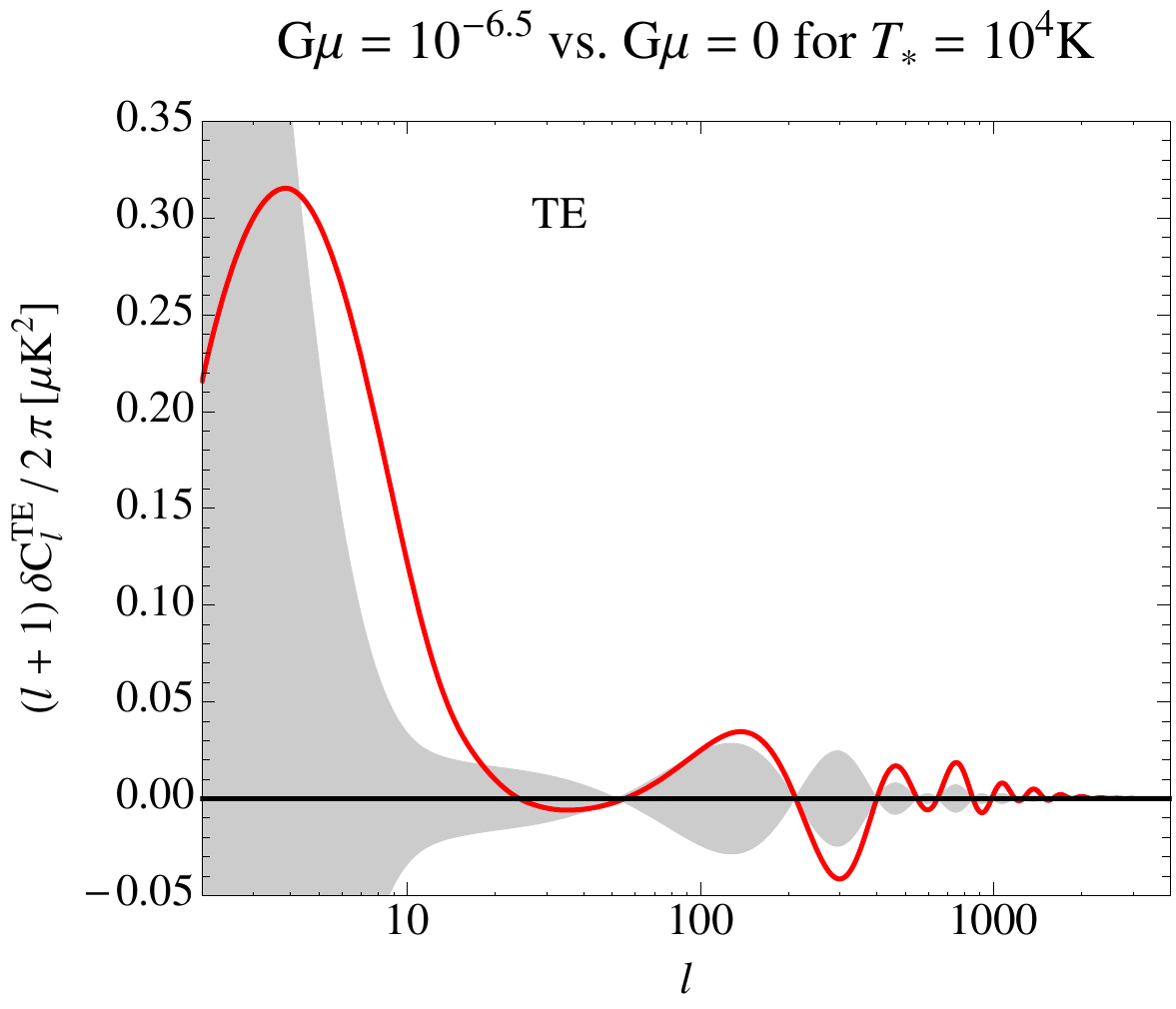}}
   	     }
   \caption{CMB TE cross-correlation with and without cosmic strings (top) and their relative difference (bottom).  We assume $T_\ast = 10^3$K (left) and $T_\ast = 10^4$K (right) for the indicated tensions vs. pure Sheth-Tormen.  The shaded region between $\pm\sqrt{2/(2l+1)}$ represents the $G\mu = 0$ cosmic variance limit for a single $C_l$.} 
   \label{fig:CTE}
\end{figure}

\section{Discussion}\label{sec:discussion}

Loops of cosmic string can seed relatively massive halos at very early
times.  These halos can become sites of early star formation and can
influence the ionization history of the universe.  Figures \ref{fig:CTT}--\ref{fig:CTE} illustrate that for $G\mu \gtrsim 10^{-7}$ the resulting effect on the CMB temperature and polarization power spectra can be significant and is likely to be detectable by Planck.  Given the present uncertainties in the physics of reionization, we did not attempt to translate our analysis into a firm bound on $G\mu$.

As we mentioned in the introduction, values of $G\mu\gtrsim 10^{-7}$ are in conflict with the
bound $G\mu \lesssim 4\times 10^{-9}$, obtained from
millisecond pulsar measurements in Refs.~\cite{vanHaasteren:2011ni,Demorest:2012bv}.
This bound, however, does not account for the fact that a significant
fraction of the string network energy goes into the kinetic energy of
loops, which is then redshifted.  It also assumes that nearly all
loops develop cusps, while examination of loops produced in the recent
simulations suggests that cusps are rather rare.  Accounting
for these effects is likely to relax the pulsar bound by at least an order of
magnitude.

We note that the current census of observed star formation at
redshifts 6--10 falls short of accounting for the minimum production
rate of UV photons that is required for ionizing the universe \cite{Bouwens,Ellis}.  
A string-seeded scenario
might be called for if future observations would reveal an abundant
population of galaxies at redshifts $z>10$.  
In the
standard cosmological model for structure formation, the comoving star
formation rate is expected to decline sharply at increasing redshifts
beyond $z\sim 10$, since the massive halos in which gas can cool and
fragment into stars become progressively rarer on the exponential tail
of the Gaussian density field. 
The cosmic string scenario provides a natural way to account for
additional star formation at high redshifts, $z\gtrsim 10$.

Early star formation from cosmic strings may help explain the observed
metallicity of the intergalactic medium (IGM) at redshifts $z \gtrsim 2$.  In particular, the observed metallicity floor
of ${\rm C_{~IV}}$ at redshifts as high as $z \sim 6$ \cite{Pettini:2003cc} 
cannot be explained with the standard star
formation history in which the filling fraction of enriched IGM is very
small at these redshifts \cite{Furlanetto:2002kr}.

Additional implications of early star formation in the cosmic string
scenario involve the existence of an abundant population of gamma-ray
bursts at unusually high redshifts,\footnote{Currently, the earliest
observed GRB is at $z \sim 9.4$ \cite{Cucchiara:2011pj}, which is
within the range expected from the standard cosmological model.} as
well as early formation of massive galaxies, which could be detected by
future infrared telescopes such as
JWST\footnote{http://www.jwst.nasa.gov/}, and early supermassive
black-holes\footnote{The possibility of early super-Eddington growth
of black holes in high redshift galaxies is reviewed in
Ref.~\cite{Wyithe:2011ap}.}, which could be detected by future X-ray
telescopes such as
IXO/Athena\footnote{http://sci.esa.int/ixo}
or gravitational wave observatories such as
LISA.\footnote{http://sci.esa.int/lisa}

Early structure formation around cosmic string loops could also be 
tested by future 21-cm observations \cite{Pritchard:2011xb}.
For example, ``global'' 21-cm measurements, which give the sky-averaged
spectrum of the 21-cm signal, are sensitive to early production of
Ly$\alpha$ photons (which couple the 21-cm excitation temperature of
hydrogen to its kinetic temperature) at $z\lesssim 50$ and to the
subsequent production of X-rays which heat the intergalactic hydrogen.
These ingredients combine to make a distinctive absorption trough in
the global 21-cm signal, which could potentially be measured by future
experiments \cite{Pritchard:2010pa}.

In this paper we focused on ``ordinary" cosmic strings, characterized by Nambu-Goto equations of motion and reconnection probability $p_{\rm rec}=1$.  The value of $p_{\rm rec}$ could be much smaller for cosmic superstrings, resulting in a denser string network, a higher density of loops, and a larger effect on structure formation (for the same value of $G\mu$).  Superstrings can also form Y-junctions, with three strings meeting at a vertex.  The evolution of networks with junctions and/or with $p_{\rm rec} <  1$ is quantitatively different from that of ordinary strings, and reliable observational predictions would require large-scale numerical simulations of such networks.

\acknowledgments We thank Ken Olum and Xavier Siemens for helpful discussions.  AL was
supported in part by NSF grant AST-0907890 and NASA grants NNX08AL43G
and NNA09DB30A.  BS and AV were supported by NSF grant PHY-0855447.
We made use of the CAMB software package to generate the
matter power spectrum and CMB angular power spectra.

\appendix
\section{The rocket effect}
\label{sec:rocket}

The gravitational radiation from a cosmic string loop is anisotropic in general.  This 
affects the momentum of the loop.  Let the radiation be beamed in the minus $y$-direction.  The physical $y$-momentum $p = m v/\sqrt{1 - v^2}$ of the loop then obeys
\beq
\frac{dp}{dt} = - H p + \Gamma_{\rm P}G\mu^2,
\eeq
where\footnote{This value was estimated for simple Kibble-Turok \cite{Turok} loop solutions involving a few lowest harmonics. Recent (non-scaling) simulations have shown that loops chopped off the network tend to be rectangular \cite{Copi:2010jw}, which can suppress $\Gamma_{\rm P}$.} 
$\Gamma_{\rm P} \sim 10$.

If we assume the non-relativistic loop has velocity $v_{\rm eq}>0$ in the $+y$-direction at $t_{\rm eq}$, its subsequent 
velocity is given by
\beq
v(a) = \frac{v_{\rm eq}}{a} + \frac{3\Gamma_{\rm P}(G\mu)^2\Meq(a^{3/2}-a^{-1})}{5m}, 
\eeq
where $a = (t/t_{\rm eq})^{2/3}$.
The trajectory in comoving coordinates is then 
\beq
y(a) = \frac{9 \Gamma_{\rm P}(G\mu)^2\Meq t_{\rm eq}}{20m}\left(\frac{4}{\sqrt{a}}+a^{2} - 5\right) + 3 v_{\rm eq} t_{\rm eq}\left(1-\frac{1}{\sqrt{a}}\right),
\eeq
and so the matter-era velocity of the loop formed at $a_i$ in the radiation era is 
\beq
\label{eq:vloopexact}
v(a,a_i) = \alpha_v\frac{a_i}{a} + \frac{(1 + 9 a^{5/2} - 10 a_i^3)\Gamma_{\rm P}G\mu}{30 a a_i^2\alpha} \approx
 \alpha_v\frac{a_i}{a} + \frac{3 a^{3/2}\Gamma_{\rm P}G\mu}{10 a_i^2\alpha}.
\eeq

From this, we can calculate the turnaround surfaces, and find the average filament thickness by
\beq
\langle x_{\rm ta}^2\rangle = \frac{\int x_{\rm ta}^4(y)dy}{\int x_{\rm ta}^2(y)dy},
\eeq
which determines the average bead mass found in section \ref{sec:filament-collapse}.  For filaments from loops produced after $a_i \sim 0.02 \mu_{-8}$, this leads to a correction factor of order $0.8$ to the average filament radius.  The only loops whose filaments receive significant corrections from the rocket effect are those whose number-densities are highly suppressed due to loop evaporation.  
Loop evaporation becomes significant only for loops of mass below
\beq
m_\Gamma = \Gamma (G\mu)^2\Meq.
\eeq
Since no star formation results from loops of mass less than $m_\ast$ given by (\ref{eq:m-star}), we can neglect loop evaporation for
\beq
\frac{9T_\ast^2\alpha_v^2}{25(G\mu)^3\alpha m_p^2} \lesssim 1,
\eeq
i.e., 
\beq
\mu_{-8} \lesssim 80\left(\frac{T_\ast}{10^4 {\rm K}}\right)^{2/3}.
\eeq
For this reason, we can neglect loop evaporation and the rocket effect entirely.


\begin{thebibliography}{9}

\bibitem{Kibble}
  T.~W.~B.~Kibble,
  ``Topology of Cosmic Domains and Strings,''
  J.\ Phys.\ A A {\bf 9}, 1387 (1976).

\bibitem{Sarangi:2002yt}
  S.~Sarangi, S.~H.~H.~Tye,
  ``Cosmic string production towards the end of brane inflation,''
  Phys.\ Lett.\  {\bf B536}, 185-192 (2002)
  [hep-th/0204074].

\bibitem{Dvali}
  G.~Dvali and A.~Vilenkin,
  ``Formation and evolution of cosmic D strings,''
  JCAP {\bf 0403}, 010 (2004)
 [hep-th/0312007].



\bibitem{Copeland}
  E.~J.~Copeland, R.~C.~Myers and J.~Polchinski,
  ``Cosmic F and D strings,''
  JHEP {\bf 0406}, 013 (2004)
 [hep-th/0312067].



\bibitem{Vlensing}
  A.~Vilenkin,
  ``Cosmic Strings As Gravitational Lenses,''
  Astrophys.\ J.\ \ {\bf 282}, L51  (1984).
  
  \bibitem{STlensing}
  B.~Shlaer and S.~-H.~H.~Tye,
  ``Cosmic string lensing and closed time-like curves,''
  Phys.\ Rev.\ D\ {\bf 72}, 043532  (2005)
  [hep-th/0502242].
  
  \bibitem{SWlensing}
  B.~Shlaer and M.~Wyman,
  ``Cosmic superstring gravitational lensing phenomena: Predictions for networks of (p,q) strings,''
  Phys.\ Rev.\ D\ {\bf 72}, 123504  (2005)
  [hep-th/0509177].
  

\bibitem{Gott} 
  J.~R.~Gott, III,
  ``Gravitational lensing effects of vacuum strings: Exact solutions,''
  Astrophys.\ J.\  {\bf 288}, 422 (1985).

\bibitem{Kaiser} 
  N.~Kaiser and A.~Stebbins,
  ``Microwave Anisotropy Due to Cosmic Strings,''
  Nature\ {\bf 310}, 391  (1984).

\bibitem{Seljak} 
  U.~Seljak and A.~Slosar,
  ``B polarization of cosmic microwave background as a tracer of strings,''
  Phys.\ Rev.\ D {\bf 74}, 063523 (2006)

\bibitem{PogosianWyman} 
  L.~Pogosian and M.~Wyman,
  ``B-modes from cosmic strings,''
  Phys.\ Rev.\ D {\bf 77}, 083509 (2008)

\bibitem{Bevis:2007qz} 
  N.~Bevis, M.~Hindmarsh, M.~Kunz and J.~Urrestilla,
  ``CMB polarization power spectra contributions from a network of cosmic strings,''
  Phys.\ Rev.\ D {\bf 76}, 043005 (2007)


\bibitem{Vilenkin-Shellard}
  A.~Vilenkin and E.~P.~S.~Shellard,
  \emph{Cosmic Strings and Other Topological Defects},
Cambridge University Press, Cambridge, England, (1994).


\bibitem{CopelandKibble} 
  E.~J.~Copeland and T.~W.~B.~Kibble,
  ``Cosmic Strings and Superstrings,''
  Proc.\ Roy.\ Soc.\ Lond.\ A {\bf 466}, 623 (2010)


\bibitem{CPV}
  E.~J.~Copeland, L.~Pogosian and T.~Vachaspati,
  ``Seeking String Theory in the Cosmos,''
  Class.\ Quant.\ Grav.\  {\bf 28}, 204009 (2011)
  [arXiv:1105.0207 [hep-th]].

\bibitem{Zeldovich} 
  Y.~.B.~Zeldovich,
  ``Cosmological fluctuations produced near a singularity,''
  Mon.\ Not.\ Roy.\ Astron.\ Soc.\  {\bf 192}, 663 (1980).

\bibitem{AV81} 
  A.~Vilenkin,
  ``Cosmological Density Fluctuations Produced by Vacuum Strings,''
  Phys.\ Rev.\ Lett.\  {\bf 46}, 1169 (1981)
  [Erratum-ibid.\  {\bf 46}, 1496 (1981)].

\bibitem{SilkV} 
  J.~Silk and A.~Vilenkin,
  ``Cosmic Strings And Galaxy Formation,''
  Phys.\ Rev.\ Lett.\  {\bf 53}, 1700 (1984).

\bibitem{Pogosian}
  M.~Wyman, L.~Pogosian, I.~Wasserman,
  ``Bounds on cosmic strings from WMAP and SDSS,''
  Phys.\ Rev.\  {\bf D72}, 023513 (2005)
 [astro-ph/0503364].
 
 
\bibitem{Fraisse} 
  A.~A.~Fraisse, C.~Ringeval, D.~N.~Spergel and F.~R.~Bouchet,
  ``Small-Angle CMB Temperature Anisotropies Induced by Cosmic Strings,''
  Phys.\ Rev.\ D {\bf 78}, 043535 (2008)
  [arXiv:0708.1162 [astro-ph]].
 
 
\bibitem{Pogosian:2008am} 
  L.~Pogosian, S.~H.~H.~Tye, I.~Wasserman and M.~Wyman,
  ``Cosmic Strings as the Source of Small-Scale Microwave Background Anisotropy,''
  JCAP {\bf 0902}, 013 (2009)
 [arXiv:0804.0810 [astro-ph]].

\bibitem{Battye}
  R.~Battye and A.~Moss,
  ``Updated constraints on the cosmic string tension,''
  Phys.\ Rev.\ D {\bf 82}, 023521 (2010)
  [arXiv:1005.0479 [astro-ph.CO]].


\bibitem{Landriau} 
  M.~Landriau and E.~P.~S.~Shellard,
  ``Cosmic String Induced CMB Maps,''
  Phys.\ Rev.\ D {\bf 83}, 043516 (2011)

\bibitem{Dvorkin}
  C.~Dvorkin, M.~Wyman and W.~Hu,
  ``Cosmic String constraints from WMAP and SPT,''
  arXiv:1109.4947 [astro-ph.CO].
  
  
  
\bibitem{vanHaasteren:2011ni}
  R.~van Haasteren, Y.~Levin, G.~H.~Janssen, K.~Lazaridis, M.~K.~B.~W.~Stappers, G.~Desvignes, M.~B.~Purver, A.~G.~Lyne {\it et al.},
  ``Placing limits on the stochastic gravitational-wave background using European Pulsar Timing Array data,''
  [arXiv:1103.0576 [astro-ph.CO]].

\bibitem{Sanidas:2012ee} 
  S.~A.~Sanidas, R.~A.~Battye and B.~W.~Stappers,
  ``Constraints on cosmic string tension imposed by the limit on the stochastic gravitational wave background from the European Pulsar Timing Array,''
  arXiv:1201.2419 [astro-ph.CO].

\bibitem{Demorest:2012bv} 
  P.~B.~Demorest, R.~D.~Ferdman, M.~E.~Gonzalez, D.~Nice, S.~Ransom, I.~H.~Stairs, Z.~Arzoumanian and A.~Brazier {\it et al.},
  ``Limits on the Stochastic Gravitational Wave Background from the North American Nanohertz Observatory for Gravitational Waves,''
  arXiv:1201.6641 [astro-ph.CO].



\bibitem{Rees}
M.~J.~Rees, 
``Baryon concentration in string wakes at $z \gtrsim 200$: Implications for galaxy formation and large-scale structure," 
MNRAS {\bf 222}, 27 (1986).

\bibitem{Hara}
  T.~Hara, S.~Miyoshi and P.~Maehoenen,
  ``On the inhomogeneity of dark matter and luminous objects induced by infinitely long cosmic strings,''
  Astrophys.\ J.\  {\bf 412}, 22 (1993).

\bibitem{Avelino}
  P.~P.~Avelino and A.~R.~Liddle,
  ``Cosmological perturbations and the reionization epoch,''
  Mon.\ Not.\ Roy.\ Astron.\ Soc.\  {\bf 348}, 105 (2004)
  [astro-ph/0305357].

\bibitem{PV}
  L.~Pogosian, A.~Vilenkin,
  ``Early reionization by cosmic strings revisited,''
  Phys.\ Rev.\  {\bf D70}, 063523 (2004)
  [astro-ph/0405606].


\bibitem{OV}
  K.~D.~Olum, A.~Vilenkin,
  ``Reionization from cosmic string loops,''
  Phys.\ Rev.\  {\bf D74}, 063516 (2006)
  [astro-ph/0605465].


\bibitem{Hernandez:2011ym}
  O.~F.~Hernandez, Y.~Wang, R.~Brandenberger, J.~Fong,
  ``Angular 21 cm Power Spectrum of a Scaling Distribution of Cosmic String Wakes,''  
  [arXiv:1104.3337 [astro-ph.CO]].

\bibitem{Brandenberger3} 
  R.~J.~Danos, R.~H.~Brandenberger and G.~Holder,
  ``A Signature of Cosmic Strings Wakes in the CMB Polarization,''
  Phys.\ Rev.\ D {\bf 82}, 023513 (2010)


\bibitem{Khatri} 
  R.~Khatri and B.~D.~Wandelt,
  ``Cosmic (super)string constraints from 21 cm radiation,''
  Phys.\ Rev.\ Lett.\  {\bf 100}, 091302 (2008)

\bibitem{Brandenberger} 
  R.~H.~Brandenberger, R.~J.~Danos, O.~F.~Hernandez and G.~P.~Holder,
  ``The 21 cm Signature of Cosmic String Wakes,''
  JCAP {\bf 1012}, 028 (2010)

\bibitem{Brandenberger2} 
  O.~F.~Hernandez, Y.~Wang, R.~Brandenberger and J.~Fong,
  ``Angular 21 cm Power Spectrum of a Scaling Distribution of Cosmic String Wakes,''
  JCAP {\bf 1108}, 014 (2011).



\bibitem{Pogosian3} 
  A.~Berndsen, L.~Pogosian and M.~Wyman,
  ``Correlations between 21 cm Radiation and the CMB from Active Sources,''
  Mon.\ Not.\ Roy.\ Astron.\ Soc.\  {\bf 407}, 1116 (2010).


\bibitem{Ringeval1} 
  C.~Ringeval, M.~Sakellariadou and F.~Bouchet,
  ``Cosmological evolution of cosmic string loops,''
  JCAP {\bf 0702}, 023 (2007)
  [astro-ph/0511646].

\bibitem{Ringeval2} 
  L.~Lorenz, C.~Ringeval and M.~Sakellariadou,
  ``Cosmic string loop distribution on all length scales and at any redshift,''
  JCAP {\bf 1010}, 003 (2010)
  [arXiv:1006.0931 [astro-ph.CO]].


\bibitem{Bennett}
  D.~P.~Bennett, F.~R.~Bouchet,
  ``Evidence for a Scaling Solution in Cosmic String Evolution,''
  Phys.\ Rev.\ Lett.\  {\bf 60}, 257 (1988); 
  D.~P.~Bennett and F.~R.~Bouchet,
  ``Cosmic String Evolution,''
  Phys.\ Rev.\ Lett.\  {\bf 63}, 2776 (1989).
  
  
 \bibitem{Allen} 
  B.~Allen and E.~P.~S.~Shellard,
  ``Cosmic String Evolution: A Numerical Simulation,''
  Phys.\ Rev.\ Lett.\  {\bf 64}, 119 (1990).
  
\bibitem{Martins} 
  C.~J.~A.~P.~Martins and E.~P.~S.~Shellard,
  ``Fractal properties and small-scale structure of cosmic string networks,''
  Phys.\ Rev.\ D {\bf 73}, 043515 (2006)
  [astro-ph/0511792].


\bibitem{OVV}
  V.~Vanchurin, K.~D.~Olum, A.~Vilenkin,
  ``Scaling of cosmic string loops,''
  Phys.\ Rev.\  {\bf D74}, 063527 (2006)
 [gr-qc/0511159].


\bibitem{OVan}
  K.~D.~Olum, V.~Vanchurin,
  ``Cosmic string loops in the expanding Universe,''
  Phys.\ Rev.\  {\bf D75}, 063521 (2007)
  [astro-ph/0610419].

\bibitem{BOS}
  J.~J.~Blanco-Pillado, K.~D.~Olum, B.~Shlaer,
  ``Large parallel cosmic string simulations: New results on loop production,''
  Phys.\ Rev.\  {\bf D83}, 083514 (2011)
  [arXiv:1101.5173 [astro-ph.CO]].

\bibitem{Vilenkin:1981iu}
  A.~Vilenkin,
  ``Cosmological Density Fluctuations Produced by Vacuum Strings,''
  Phys.\ Rev.\ Lett.\  {\bf 46}, 1169-1172 (1981).

\bibitem{Albrecht:1984xv}
  A.~Albrecht, N.~Turok,
  ``Evolution of Cosmic Strings,''
  Phys.\ Rev.\ Lett.\  {\bf 54}, 1868-1871 (1985).

\bibitem{JJP}
  M.~G.~Jackson, N.~T.~Jones and J.~Polchinski,
  ``Collisions of cosmic F and D-strings,''
  JHEP {\bf 0510}, 013 (2005)
 [hep-th/0405229].


\bibitem{Damour}
  T.~Damour and A.~Vilenkin,
  ``Gravitational radiation from cosmic (super)strings: Bursts, stochastic background, and observational windows,''
  Phys.\ Rev.\ D {\bf 71}, 063510 (2005)
  [hep-th/0410222].
  
  
\bibitem{Sakellariadou}
  M.~Sakellariadou,
  ``A Note on the evolution of cosmic string/superstring networks,''
  JCAP {\bf 0504}, 003 (2005)
  [hep-th/0410234].

\bibitem{Shellard}
  A.~Avgoustidis and E.~P.~S.~Shellard,
  ``Effect of reconnection probability on cosmic (super)string network density,''
  Phys.\ Rev.\ D {\bf 73}, 041301 (2006)
  [astro-ph/0512582].

\bibitem{Hindmarsh:2008dw} 
  M.~Hindmarsh, S.~Stuckey and N.~Bevis,
  ``Abelian Higgs Cosmic Strings: Small Scale Structure and Loops,''
  Phys.\ Rev.\ D {\bf 79}, 123504 (2009)

\bibitem{ShellardMoore} 
  J.~N.~Moore and E.~P.~S.~Shellard,
  ``On the evolution of Abelian Higgs string networks,''
  hep-ph/9808336.

\bibitem{OlumJose} 
  K.~D.~Olum and J.~J.~Blanco-Pillado,
  ``Radiation from cosmic string standing waves,''
  Phys.\ Rev.\ Lett.\  {\bf 84}, 4288 (2000)


\bibitem{Polchinski}
  F.~Dubath, J.~Polchinski and J.~V.~Rocha,
  ``Cosmic String Loops, Large and Small,''
  Phys.\ Rev.\ D {\bf 77}, 123528 (2008)
  [arXiv:0711.0994 [astro-ph]].

\bibitem{Loeb-book}
A.~Loeb,
{\em How Did the First Stars and Galaxies Form?},
Princeton University Press, Princeton, NJ, U.S.A., (2010).

\bibitem{Stacy:2010gg}
  A.~Stacy, V.~Bromm, A.~Loeb,
  ``Effect of Streaming Motion of Baryons Relative to Dark Matter on the Formation of the First Stars,''  
  [arXiv:1011.4512 [astro-ph.CO]].


\bibitem{Tseliakhovich:2010yw}
  D.~Tseliakhovich, R.~Barkana, C.~Hirata,
  ``Suppression and Spatial Variation of Early Galaxies and Minihalos,''
  [arXiv:1012.2574 [astro-ph.CO]].



\bibitem{Eisenstein:1996pr}
  D.~J.~Eisenstein, A.~Loeb, E.~L.~Turner,
  ``Dynamical mass estimates of large scale filaments in redshift surveys,''
  Submitted to: Astrophys. J.
  [astro-ph/9605126].



\bibitem{Bertschinger}
  E.~Bertschinger,
  ``Cosmological accretion wakes,''
  Astrophys.\ J.\  {\bf 316}, 489 (1987).

\bibitem{Quillen:2010yc}
  A.~C.~Quillen, J.~Comparetta,
  ``Jeans Instability of Palomar 5's Tidal Tail,''
  [arXiv:1002.4870 [astro-ph.CO]].

  
  


  
\bibitem{Sheth:1999mn}
  R.~K.~Sheth, G.~Tormen,
  ``Large scale bias and the peak background split,''
  Mon.\ Not.\ Roy.\ Astron.\ Soc.\  {\bf 308}, 119 (1999)
  [astro-ph/9901122].
  
\bibitem{Komatsu:2010fb} 
  E.~Komatsu {\it et al.}  [WMAP Collaboration],
  ``Seven-Year Wilkinson Microwave Anisotropy Probe (WMAP) Observations: Cosmological Interpretation,''
  Astrophys.\ J.\ Suppl.\  {\bf 192}, 18 (2011)
 [arXiv:1001.4538 [astro-ph.CO]].
 
 
  
 
\bibitem{BL01} 
  R.~Barkana and A.~Loeb,
  ``In the beginning: The First sources of light and the reionization of the Universe,''
  Phys.\ Rept.\  {\bf 349}, 125 (2001)
 [astro-ph/0010468].

\bibitem{WL03} 
  J.~S.~B.~Wyithe and A.~Loeb,
  ``Reionization of hydrogen and helium by early stars and quasars,''
  Astrophys.\ J.\  {\bf 586}, 693 (2003)
 [astro-ph/0209056].

 \bibitem{BKL} 
  V.~Bromm, R.~P.~Kudritzki and A.~Loeb,
  ``Generic spectrum and ionization efficiency of a heavy initial mass function for the first stars,''
  Astrophys.\ J.\  {\bf 552}, 464 (2001)
  [astro-ph/0007248].

  
  \bibitem{Trac} 
  H.~Trac and R.~Cen,
  ``Radiative transfer simulations of cosmic reionization. 1. Methodology and initial results,''
  Astrophys.\ J.\ {\bf 671}, 1 (2007)
  [astro-ph/0612406].

  
\bibitem{Lewis:1999bs} 
  A.~Lewis, A.~Challinor and A.~Lasenby,
  ``Efficient computation of CMB anisotropies in closed FRW models,''
  Astrophys.\ J.\  {\bf 538}, 473 (2000)
  [astro-ph/9911177].
 
\bibitem{:2006uk} 
  [Planck Collaboration],
  ``The Scientific programme of planck,''
  astro-ph/0604069.


\bibitem{Bouwens} 
  R.~J.~Bouwens, G.~D.~Illingworth, P.~A.~Oesch, M.~Trenti, I.~Labbe, M.~Franx, M.~Stiavelli and C.~M.~Carollo {\it et al.},
  ``Lower-Luminosity Galaxies could reionize the Universe: Very Steep Faint-End Slopes to the UV Luminosity Functions at z $\geq$ 5-8 from the HUDF09 WFC3/IR Observations,''
  arXiv:1105.2038 [astro-ph.CO].


\bibitem{Ellis} 
  B.~E.~Robertson, R.~S.~Ellis, J.~S.~Dunlop, R.~J.~McLure and D.~P.~Stark,
  ``Early star-forming galaxies and the reionization of the Universe,''
  Nature {\bf 468}, 49 (2010)
  [arXiv:1011.0727 [astro-ph.CO]]



\bibitem{Pettini:2003cc} 
  M.~Pettini, P.~Madau, M.~Bolte, J.~X.~Prochaska, S.~L.~Ellison and X.~Fan,
  ``The ${\rm C_{~IV}}$ mass density of the universe at redshift 5,''
  Astrophys.\ J.\  {\bf 594}, 695 (2003)
  [astro-ph/0305413].


\bibitem{Furlanetto:2002kr}
  S.~R.~Furlanetto and A.~Loeb,
  ``Metal Absorption Lines as Probes of the Intergalactic Medium Prior to the
  Reionization Epoch,''
  Astrophys.\ J.\  {\bf 588}, 18 (2003)
  [arXiv:astro-ph/0211496].


\bibitem{Cucchiara:2011pj} 
  A.~Cucchiara, A.~J.~Levan, D.~B.~Fox, N.~R.~Tanvir, T.~N.~Ukwatta, E.~Berger, T.~Kruhler and A.~K.~Yoldas {\it et al.},
  ``A Photometric Redshift of z $\sim$ 9.4 for GRB 090429B,''
  Astrophys.\ J.\  {\bf 736}, 7 (2011)
  [arXiv:1105.4915 [astro-ph.CO]].

\bibitem{Pritchard:2011xb} 
  J.~R.~Pritchard and A.~Loeb,
  ``21-cm cosmology,''
  arXiv:1109.6012 [astro-ph.CO].



\bibitem{Pritchard:2010pa} 
  J.~R.~Pritchard and A.~Loeb,
  ``Constraining the unexplored period between the dark ages and reionization with observations of the global 21 cm signal,''
  Phys.\ Rev.\ D {\bf 82}, 023006 (2010)
  [arXiv:1005.4057 [astro-ph.CO]].


\bibitem{Wyithe:2011ap} 
  S.~Wyithe and A.~Loeb,
  ``Photon Trapping Enables Super-Eddington Growth of Black-Hole Seeds in Galaxies at High Redshift,''
  arXiv:1111.5424 [astro-ph.CO].

\bibitem{Turok}
 T.~W.~B.~Kibble and N.~Turok,
  ``Selfintersection of Cosmic Strings,''
  Phys.\ Lett.\ B {\bf 116}, 141 (1982).
  
\bibitem{Copi:2010jw}
  C.~J.~Copi, T.~Vachaspati,
  ``Shape of Cosmic String Loops,''
  Phys.\ Rev.\  {\bf D83}, 023529 (2011)
  [arXiv:1010.4030 [hep-th]].


\end{thebibliography}
\end{document}